\documentclass[3p,english,number,times]{elsarticle}
\makeatletter
\def\ps@pprintTitle{%
 \let\@oddhead\@empty
 \let\@evenhead\@empty
 \def\@oddfoot{}%
 \let\@evenfoot\@oddfoot}
\makeatother
\biboptions{sort&compress}
\usepackage{rotating,graphicx,epsfig,epstopdf,float}
\usepackage{amsmath,amssymb}
\usepackage{makeidx,url}
\usepackage{array,xspace}
\usepackage{hyperref,booktabs}

\usepackage[symbol]{footmisc}

\hypersetup{
 pdftitle={A search for the LHCb Charmed 'Pentaquark' using Photoproduction of
 J/psi at Threshold in Hall C at Jefferson Lab},%
 pdfauthor={Z.-E.~Meziani, S.~Joosten, M. Paolone, E. Chudakov, M. Jones},%
 pdfsubject={},%
 pdfkeywords={Quarkonium Production, Fixed-Target experiment, JLab},%
 pdfstartview={},%
 bookmarksopen=true, breaklinks=true, debug=true, %
 colorlinks=true, linkcolor=red, citecolor=blue, urlcolor=blue
}

\newcommand{\jpsi}{\ensuremath{J/\psi}\xspace}
\newcommand{\pc}{\ensuremath{P_c(4450)}\xspace}
\newcommand{\pcwide}{\ensuremath{P_c(4380)}\xspace}

\begin{document}

\author[temple]{Z.-E.~Meziani\corref{contact}}
\ead{meziani@temple.edu}
\author[temple]{S.~Joosten\corref{cospoke}}
\ead{sylvester.joosten@temple.edu}
\author[temple]{M.~Paolone\corref{cospoke}}
\ead{michael.paolone@temple.edu}
\author[JLab]{E.~Chudakov\corref{cospoke}}
\ead{gen@jlab.org}
\author[JLab]{M.~Jones\corref{cospoke}}
\ead{jones@jlab.org}
\author[miss]{K.~Adhikari}
\author[csu]{K.~Aniol}
\author[anl]{W.~Armstrong}
\author[anl]{J.~Arrington}
\author[yerevan]{A.~Asaturyan}
\author[temple]{H.~Atac}
\author[seoul]{S.~Bae}
\author[miss]{H.~Bhatt}
\author[miss]{D.~Bhetuwal}
\author[JLab]{J.-P.~Chen}
\author[imp]{X.~Chen}
\author[seoul]{H.~Choi}
\author[seoul]{S.~Choi}
\author[JLab]{M.~Diefenthaler}
\author[miss]{J.~Dunne}
\author[orsay]{R.~Dupr\'e}
\author[temple]{B.~Duran}
\author[miss]{D.~Dutta}
\author[miss]{L.~El-Fassi}
\author[imp]{Q.~Fu}
\author[duke]{H.~Gao}
\author[seoul]{H.~Go}
\author[duke]{C.~Gu}
\author[seoul]{J.~Ha}
\author[anl]{K.~Hafidi}
\author[JLab]{O.~Hansen}
\author[anl]{M.~Hattawy}
\author[JLab]{D.~Higinbotham}
\author[regina]{G.~M.~Huber}
\author[fiu]{P.~Markowitz}
\author[JLab]{D.~Meekins}
\author[yerevan]{H.~Mkrtchyan}
\author[bnl]{R.~Nouicer}
\author[regina]{W.~Li}
\author[duke]{X.~Li}
\author[duke]{T.~Liu}
\author[duke]{C.~Peng}
\author[JLab]{L.~Pentchev}
\author[JLab]{E.~Pooser}
\author[temple]{M.~Rehfuss}
\author[temple]{N.~Sparveris}
\author[yerevan]{V.~Tadevosyan}
\author[imp]{R.~Wang}
\author[odu]{F.~R.~Wesselmann}
\author[JLab]{S.~Wood}
\author[duke]{W.~Xiong}
\author[duke]{X.~Yan}
\author[miss]{L.~Ye}
\author[anl]{Z.~Ye}
\author[regina]{A.~Zafar}
\author[imp]{Y.~Zhang}
\author[imp]{F.~Zhao}
\author[duke]{Z.~Zhao}
\author[yerevan]{S.~Zhamkochyan}
\address[yerevan]{Alikhanyan National Science Laboratory, Yerevan, Armenia}
\address[anl]{Argonne National Laboratory, Chicago, IL}
\address[bnl]{Brookhaven National Laboratory, Upton, NY}
\address[csu]{California State University, Los Angeles, CA}
\address[duke]{Duke University, Durham, NC}
\address[fiu]{Florida International University, Miami, FL}
\address[miss]{Mississippi State University, Starkville, MS}
\address[odu]{Old Dominion University, Norfolk, VA}
\address[imp]{Institute of Modern Physics, Chinese Academy of Sciences, Lanzhou,
China}
\address[orsay]{Institut de Physique Nucléaire d'Orsay, Orsay, France}
\address[regina]{University of Regina, SK, Canada}
\address[seoul]{Seoul National University, Seoul, Korea}
\address[JLab]{Thomas Jefferson National Accelerator Facility, Newport News, VA}
\address[temple]{Temple University, Philadelphia, PA}

\cortext[contact]{Co-Spokesperson/Contact}
\cortext[cospoke]{Co-Spokesperson}

\title{\large A Search for the LHCb Charmed `Pentaquark' using Photo-Production
of $J/\psi$ at Threshold in Hall C at Jefferson Lab}
\date{August 31, 2016}

\begin{abstract} 
We propose\footnote[2]{
  This document is an updated version of the original proposal PR12-16-007, which
  was approved with an `A' rating and a `high-impact' label by the Jefferson Lab
  PAC 44 in July 2016. The experiment was awarded 11 days of beam time.}
to measure the photo-production cross section of
\jpsi near threshold, in search of the recently observed LHCb hidden-charm
resonances $P_c$(4380) and $P_c$(4450) consistent with `pentaquarks'. 
The observation of these resonances in photo-production will provide strong
evidence of the true resonance nature of the LHCb states, distinguishing them
from kinematic enhancements.
A bremsstrahlung photon beam produced with an $11\,\text{GeV}$ electron beam at
CEBAF covers the energy range of \jpsi production from the threshold
photo-production energy of $8.2\,\text{GeV}$, to an energy beyond the presumed
\pc resonance.
The experiment will be carried out in Hall C at Jefferson Lab, using a 
$50\,\mu\text{A}$ electron beam incident on a 9\% copper radiator. 
The resulting photon beam passes through a $15\,\text{cm}$ liquid hydrogen target,
producing \jpsi mesons through a diffractive process in the $t$-channel,
or through a resonant process in the $s$- and $u$-channel.
The decay $e^+e^-$ pair of the \jpsi will be detected in coincidence using the
two high-momentum spectrometers of Hall C.
The spectrometer settings have been optimized to distinguish the resonant $s$- and $u$-channel
production from the diffractive $t$-channel \jpsi production.
The $s$- and $u$-channel
production of the charmed 5-quark resonance dominates the $t$-distribution at large $t$. 
The momentum and angular resolution of the spectrometers is sufficient
to observe a clear resonance enhancement in the total cross section and $t$-distribution. 
We request a total of 11 days of beam time with 9 days to
carry the main experiment and 2 days to acquire the needed
$t$-channel elastic $J/\psi$ production data for a calibration measurement.
This calibration measurement in itself will greatly enhance our knowledge of 
$t$-channel elastic \jpsi production near threshold.

\end{abstract}

\maketitle

\tableofcontents 


\section{Introduction and motivation}
\label{sec:intro}
Photo-production of $J/\psi$ on a nucleon very close to threshold is an important
subject in the field of non-perturbative QCD in its own right~\cite{Brodsky:2015zxu} and is already planned to be investigated at Jefferson Lab as the 12 GeV upgrade of CEBAF is
completed~\cite{CLAS12-tcs:proposal,SoLIDjpsi:proposal}. Oddly enough the
potential of discovery of hidden charm baryon resonances via photo-production was
discussed in 2014~\cite{Huang:2013mua} inspired in part by the SoLID-$J/\psi$
approved proposal at Jefferson Lab~\cite{SoLIDjpsi:proposal}. However, CERN's recent
experimental discovery~\cite{cernpr:2016} has spurred a new excitement and a
sense of urgency to carry out measurements of photo-production at threshold in a timely manner. 

Less than a year ago, on July 14, 2015, a press release from the CERN press
office announced  the observation of exotic pentaquark
particles~\cite{cernpr:2016} just a day after the manuscript describing the
discovery was posted on the arXiv.org~\cite{Sheldon:2016} website by the
collaboration. A month  later, on August 12, 2015 the announcement was followed
by the publication of the manuscript describing the discovery in Physical Review
Letters~\cite{Aaij:2015tga}. This announcement was received with both excitement
and a healthy dose of skepticism due to the early saga of `pentaquarks,' in the beginning of the new millennium, which proved
inconclusive. Unlike these earlier announced pentaquarks, which consisted of four light quarks and one strange quark, the resonant state observed by the LHCb includes two heavy quarks, namely charm and anti-charm quarks and thus must be different in nature.

Subsequent to the announcement a series of theoretical
papers~\cite{Liu:2016fe,Karliner:2015ina,Chen:2015loa,Eides:2015dtr,Wang:2015jsa,Karliner:2015voa,Kubarovsky:2015aaa,Guo:2015umn}
appeared
in the literature with possible interpretations of the observed resonance. A
range of explanations was invoked, from a possible true pentaquark resonant
state to a kinematic enhancement like those observed in other experiments close
to kinematic thresholds~\cite{Bai:2003sw}, such as a bound state of charmonium$
(2S)$ and the proton~\cite{Eides:2015dtr}, or a molecule composed of $\Sigma_c$
and $\bar D^*~$\cite{Lu:2016nnt,Huang:2016tcr}. But without further experimental
measurements it is not clear whether the formed exotic resonance can be
unambiguously identified as a resonance. Some authors suggested that effects of
final state interactions are responsible for the LHCb observed rate
enhancements~\cite{Guo:2015umn}. While the interest of the theory community has
produced more than 200 citations up to date, LHCb is the only experiment that
has observed these states. The hadronic physics community is eager to see these possible resonant states confirmed in more than one experiment and  proceed with a detailed investigation of the quantum numbers of such states. 

In summary, to resolve the true nature of the $P_c^+(4380)$ and $P_c^+(4450)$
states it is proposed to study these pentaquark candidates in direct
photo-production of $J/\psi$ on the proton and provide not only further evidence
of their existence but also investigate their spin and parity, as noted in
several papers, such
as~\cite{Wang:2015jsa,Karliner:2015voa,Kubarovsky:2015aaa,Blin:2016dlf}.
This proposal is  more specifically about a direct search of the higher mass
narrow width $P_c^+(4450)$ and follows Wang et al.~\cite{Wang:2015jsa} using the
different spins and parity described in the paper but with the less optimistic
assumptions about the coupling to the resonant states during our complete
simulations, namely a 5\% coupling.

We believe that the results from this search at Jefferson Lab will have a high
impact on the broader physics community.
 
 \subsection{Present data status}

\begin{figure}[!ht]
\begin{center}
\includegraphics[width=0.66\textwidth]{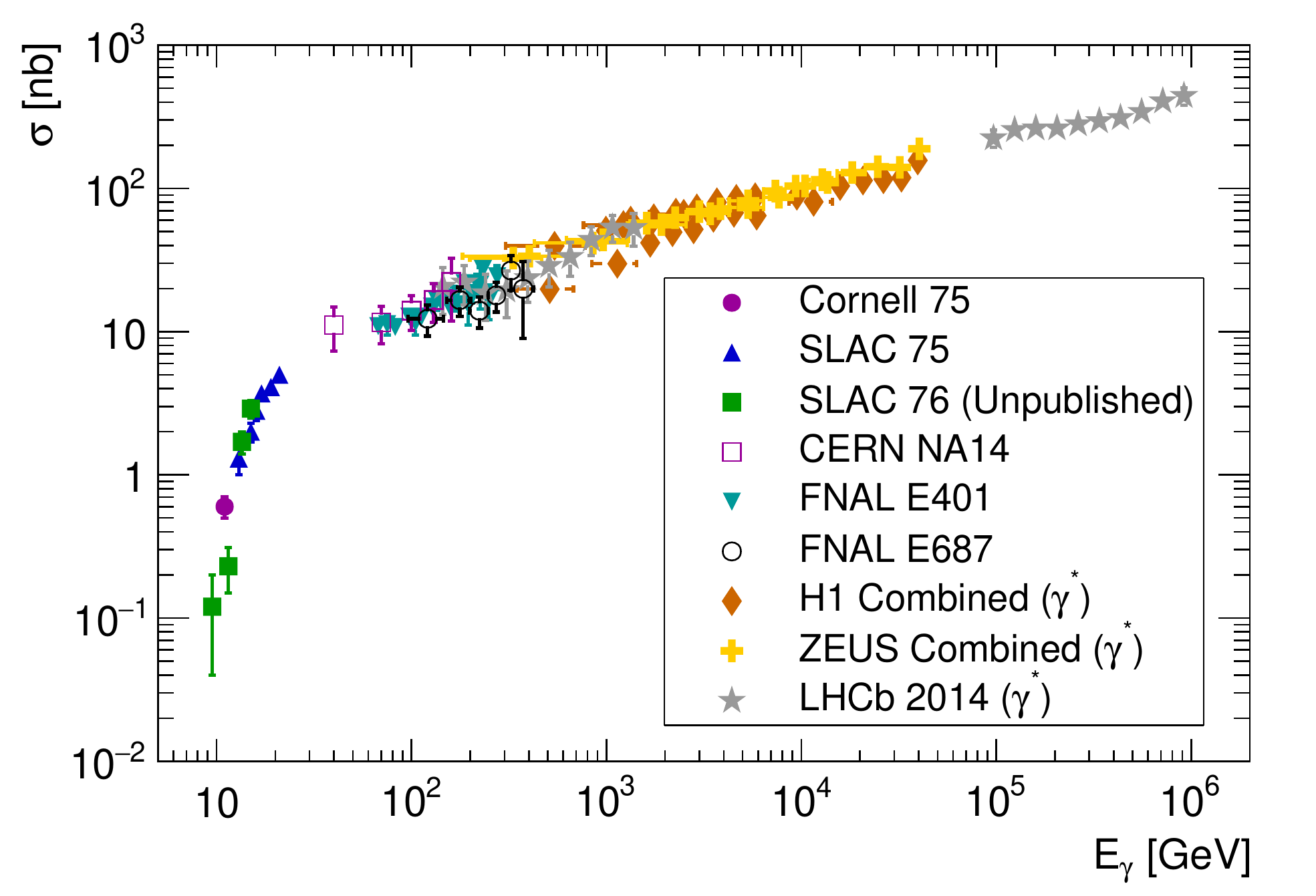}
\includegraphics[width=0.33\textwidth]{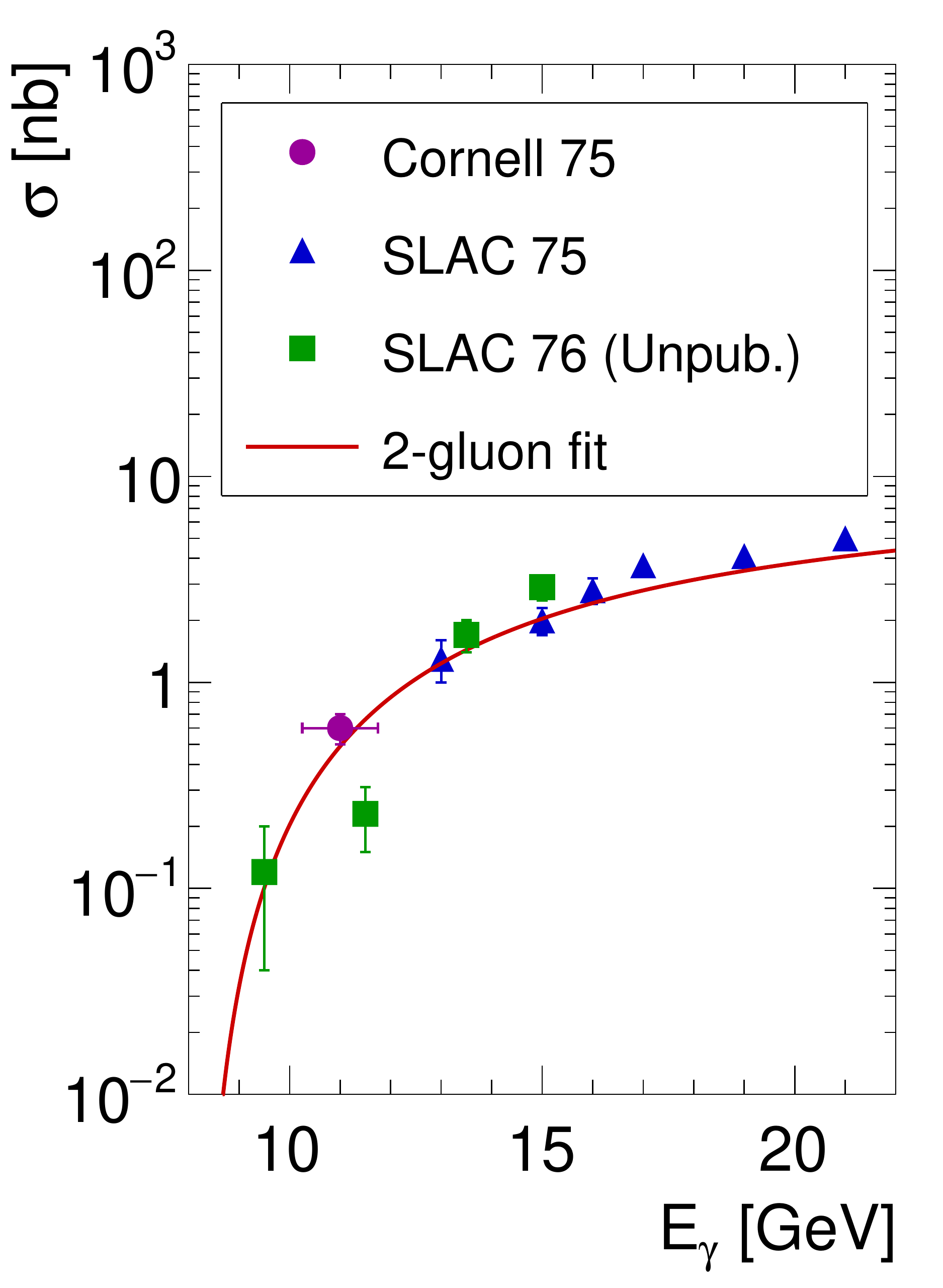}
\caption{Compilation of world data for the electro- and photo-production of
elastic \jpsi. Shown in the left panel are Cornell data from \cite{Gittelman:1975ix}, SLAC data from
\cite{Camerini:1975cy,Anderson:1976sd}, CERN NA14 data from \cite{Barate:1986fq},
FNAL data from \cite{Binkley:1981kv,Frabetti:1993ux}, H1 data from
\cite{Adloff:2000vm,Alexa:2013xxa}, ZEUS data from \cite{Chekanov:2002xi} and LHCb data from 
\cite{Aaij:2013jxj}. Legend in the figure with $\gamma^*$ refer to
electro-production data  and thus an effective photon energy defined by $E_{\gamma}^{eff} = (W_{\gamma p}^2-M_p^2 )/2M_p$ was used. The right panel zooms on the region of
interest near the \jpsi production threshold region. The red curve on the right figure is the result of a 2-gluon fit. }
\label{fig:xsection}
\end{center}
\end{figure}

The photo-production of $J/\psi$ has been measured in many experiments at high
invariant mass of the photon-proton system ($W_{\gamma-p}$ at HERA~\cite{Chekanov:2002xi,Alexa:2013xxa}, and more recently at LHCb~\cite{Aaij:2013jxj} (see right Fig.~\ref{fig:xsection}). The total elastic $J/\psi$ production at high photon-nucleon invariant mass  $W_{\gamma p}$ is well described by the $t$-channel
exchange of a colorless object between the photon and the
proton~\cite{frankfurt:2002}, in this case two-gluon exchange. The differential
cross section in the proton momentum transfer variable $t$ is usually described
by $d\sigma/dt \propto e^{bt}$ with a value of $b$ that depends on 
 $W_{\gamma p}$. As $W_{\gamma p}$ decreases towards the
threshold region of $J/\psi$ production, the mechanism is described by a Pomeron
exchange or two-gluon exchange~\cite{Brodsky:2000zc} or perhaps a more
complicated multi-gluon exchange carrying the non-perturbative information of
the gluonic fields in the nucleon. The new LHCb resonance happens to be in this
threshold region of invariant mass, a region that has been poorly explored in
modern times. It is worth pointing out that the few measurements of this region
occurred in the 1970s at Cornell and SLAC and in the 80s at Fermilab ( see left Fig.~\ref{fig:xsection}). In those experiments, issues of unambiguously defining the elastic process of $J/\psi$
production were hampered in some cases by the use of nuclear targets, detector
resolution and the detection of one lepton only in the case of the $J/\psi$ pair decay.

In Hall C at Jefferson Lab, a photo-production experiment (E03-008) was
performed in the {\it subthreshold} regime, but unfortunately no signal was
observed after one week of beam scattering off a $^{12}$C
target~\cite{Bosted:2008mn}. The experiment used a bremsstrahlung beam produced in a copper radiator by the 6 GeV incident electron beam at CEBAF. The pair of spectrometers (HMS and SOS) of Hall C  were used to detect the pair of leptons resulting from the decay of the
$J/\psi$. This experiment allowed an upper limit to be set on the cross
section which was found to be consistent with the quasi-free production.  
More recently a proposal~\cite{Chudakov:2007} for the 12 GeV upgrade of
Hall C was considered by the PAC and conditionally approved. The authors
proposed again the use of bremsstrahlung photon beam created in a radiator to
look at the photo-production at threshold in a series of nuclei. The physics goal
was to measure the photo-production cross section in order to investigate the
A-dependence of the propagation of the $J/\psi$ in the nuclear medium as well as
extract the $J/\psi-N$ interaction. In the latter proposal, the $J/\psi$ decay
pair was to be detected by the HMS and SHMS similar to what is proposed here, however the optimization of the spectrometer settings was related to enhancing the rate of pairs detected from $J/\psi$ decays in primarily diffractive $J/\psi$ production off nuclei, no resonant production was considered. 

In summary, the near threshold region of elastic $J/\psi$ production has yet to
be fully explored in the context of understanding the non-perturbative gluonic
$J/\psi$-nucleon interaction. At Jefferson Lab there are approved proposals to
measure this region using the CLAS12 detector in Hall B~\cite{CLAS12-tcs:proposal} and the SoLID
detector in Hall A~\cite{SoLIDjpsi:proposal}. In this proposal our focus is to
confirm the observation of LHCb through a resonant production of the $P_c(4450)$
in the $s$- and $u$-channel.

\section{The proposed measurement in Hall C at Jefferson Lab}

We propose to measure the elastic $J/\psi$ photo-production cross section as a
function of $t$ and photon energy $E_{\gamma}$ in the near threshold region in
Hall C. A bremsstrahlung photon beam will be created using a 9\% copper radiator
in front of a liquid hydrogen target, similar to the E-05-101
experiment~\cite{Fanelli:2015eoa}. The optimal placement of the radiator will be
chosen to account for the closer proximity of the flow diverters to the beam.

Both high momentum spectrometers of Hall C
along with their associated detectors will be used to detect the di-lepton pair
decay, namely e$^+$e$^-$. The photon beam mixed with the primary electron beam
will strike a 15 cm liquid hydrogen target. The electron-positron decay pair
will be detected in coincidence between the high momentum spectrometer (HMS) set
for electron detection and the super-high momentum spectrometer (SHMS) set for
positron detection. Both spectrometer arms will be used in their standard
configuration.
  
The proposed measurement is designed to search for the highest mass narrow
exotic resonant state discovered at LHCb, namely the $P_c(4450)$. The
spectrometer settings (shown in Tab.~\ref{table:kin}) are optimized to be most
sensitive to the possible resonant production of $P_c$(4450) in the $s$- and
$u$-channel.  The two spectrometers will detect the $J/\psi$ decay into
e$^+$e$^-$ from either the diffractive channel or resonant $P_c$ channel
production. However, we will take advantage of the different $t$-dependence of
the two processes to optimize the spectrometers' angle and momentum settings to
enhance the $P_c(4450)$ signal relative to that of the $t$-channel production.

\subsection{The experiment in Hall C}

\begin{figure}
\begin{center}
\includegraphics[scale=1.0,width=\textwidth]{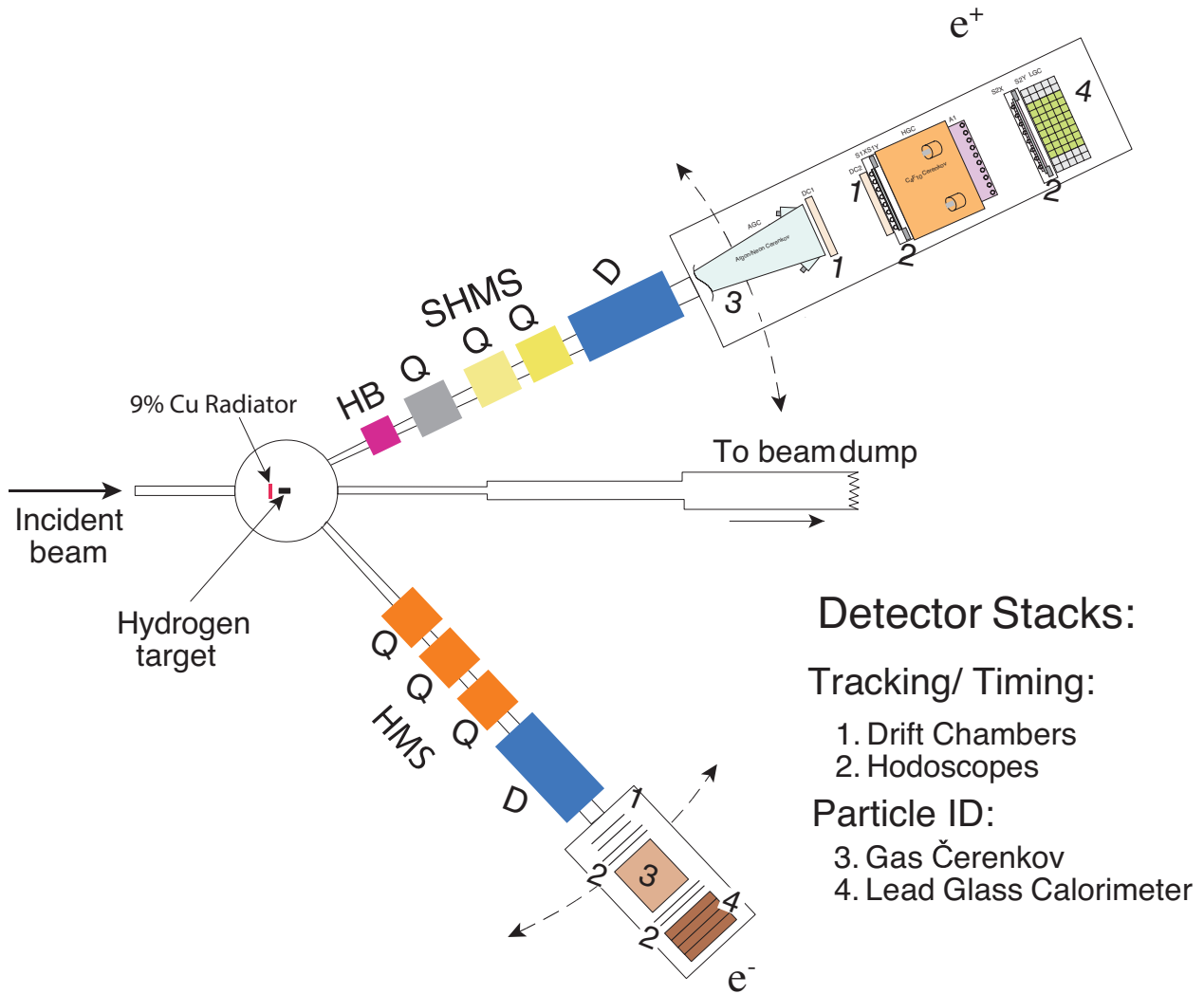}
\caption{The experimental layout of the HMS (for $e^-$ detection ) and SHMS (for
e$^+$ detection) and associated  detectors combined with a liquid 15 cm hydrogen
target and a 9\% copper radiator. We will detect the $J/\psi$ decay  e$^+$e$^-$
pair in coincidence between the two spectrometers. From the scattering angle and
momentum determination of the lepton pairs, we are able to reconstruct the
invariant mass of the $J/\psi$ as well as its three-momentum. From this
information the four-momentum transfer to the proton $t$ is calculated and the
real photon energy $E_{\gamma}$ is determined.}
\label{exp:layout}
\end{center}
\end{figure}

The layout of the proposed experiment is shown in Fig.~\ref{exp:layout} where
the HMS is set at an angle of 34$^{\circ}$ to detect the electrons of the
e$^+$e$^-$ decay pairs while the SHMS is set at angle of 13$^{\circ}$ to detect
the corresponding positrons. This configuration has been optimized to reduce the
accidental coincidences between the two spectrometers as well as minimize the
absolute background in each spectrometer. Part of the momentum acceptance of the
spectrometer will allow for the detection of the Bethe-Heitler process in a
kinematic region forbidden to the diffractive or resonant production of
$J/\psi$. The HMS is chosen to detect electrons, because the inclusive inelastic electron
scattering cross section drops rapidly with increasing scattering angle. On the
other hand the SHMS would have a very large rate of inclusive electron
scattering if it were to be used to detect electrons at the small angle setting
of 13$^{\circ}$. It is thus run in positive polarity to detect positrons, but it
will also accept positive pions and protons.

Each spectrometer has a similar set of standard detectors to identify
electrons/positrons and reject charged pions and protons. In each case the
momentum of the particles is provided through tracking by a set of drift
chambers, and electron identification is ensured by a light gas \v Cerenkov
counter and an electromagnetic calorimeter. The trigger in each spectrometer is
defined by a coincidence between a set of 2 hodoscope scintillator planes along
the path of the particles.  These configurations offer an electron or positron
detection efficiency greater than 98\% and a pion rejection factor of about a
thousand. The timing resolution online will be defined by the time coincidence
between the hodoscopes in each spectrometer first and then a time coincidence
between both spectrometers. We expect it to be on the range of few nanoseconds,
however improvements will be possible when the tracking information is corrected
for offline. A gate of about 50 ns will be used between the two spectrometers.

\begin{table}[H]
\caption{Kinematic setting of the HMS and SHMS spectrometers to measure in
coincidence the decay-pair of the \jpsi. The main spectrometer setting (1) is
optimized to measure the $P_c(4450)$ with minimal $t$-channel background production, while the
additional setting (2) is chosen to allow for a precise determination of the
$t$-channel background left of the $P_c(4450)$ resonance.}
\begin{center}
\begin{tabular}{c|cc|cc|cc}
      &\multicolumn{2}{c|}{HMS} & \multicolumn{2}{c|}{SHMS} &
        \multicolumn{2}{c}{Acceptance}\\
      \hline
      &$p$  & $\theta$ & $p$  & $\theta$ & t-channel & $P_c(4450)$\\
      Setting&GeV/$c$& &GeV/$c$&&\%&\%\\
      \hline
      \#1 & 3.25  &  34.5$^\circ$ & 4.5 & 13.0$^\circ$  & 0.0004 & 0.003\\ 
      \#2 & 4.75  &  20.0$^\circ$ & 4.25 & 20.0$^\circ$  & 0.01 & 0.003 
\end{tabular}
\end{center}
\label{table:kin}
\end{table}%

We point out that a lower electron beam energy of 10.7 GeV, rather than 11.0
GeV, with the same spectrometer settings, will result in a more
suppressed $t$-channel production, while it should not affect the $s$ and $u$
channel $P_c(4450)$ production.

\subsection{Kinematics}

\begin{figure}[htb]
\begin{center}
\includegraphics[scale=0.6,angle=-90,width=0.75\textwidth]{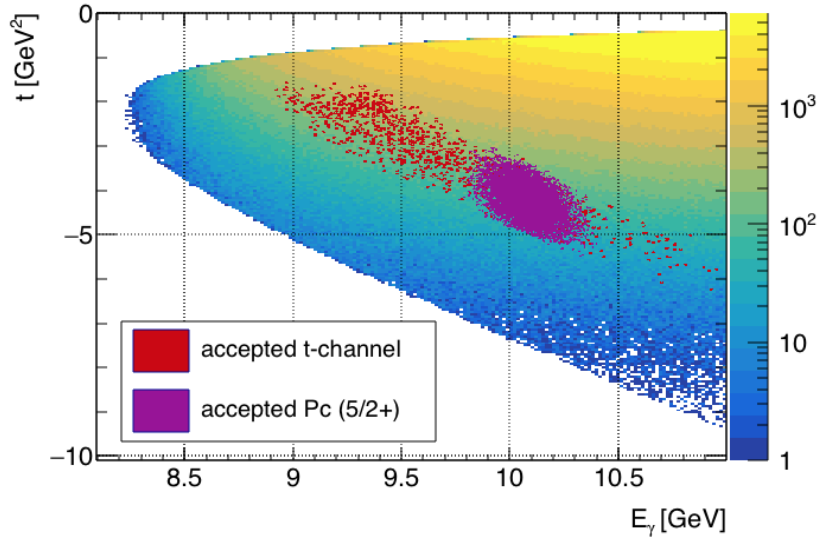}
\caption{Full phase space of the $t$ channel elastic $J/\psi$ production with the acceptance rate shown. The variable $t$ versus incoming photon energy $E_{\gamma}$ is plotted. The kinematic setting \#1 for this experiment with the accepted $t$ channel events (red) and the $P_c(4450)- 5/2+$ (purple)  resonant state events. 
}
\label{fig:kin}
\end{center}
\end{figure}

The kinematics were optimized using a full simulation of the experiment and
focused on enhancing the resonant production of $J/\psi$ through the $P_c(4450)$
relative to the diffractive production. This is done by taking advantage of the
$t$-dependence of the diffractive production since the latter is suppressed at
large values of $t$ while the resonant production in the $s$-channel of $P_c$ is
rather flat across the same $t$ range. The spectrometer settings are chosen to
take advantage of this difference in $t$-dependence. In Tab.~\ref{table:kin} we
list the spectrometers' momentum and angle settings converged upon and the
resulting acceptance for a coincident detection of the di-lepton pair. Also
shown is the additional spectrometer setting needed for a precise determination
of the $t$-channel background left of the \pc peak.

\begin{figure}[!ht]
\begin{center}
\includegraphics[scale=1.0,angle=-90,width=0.47\textwidth]{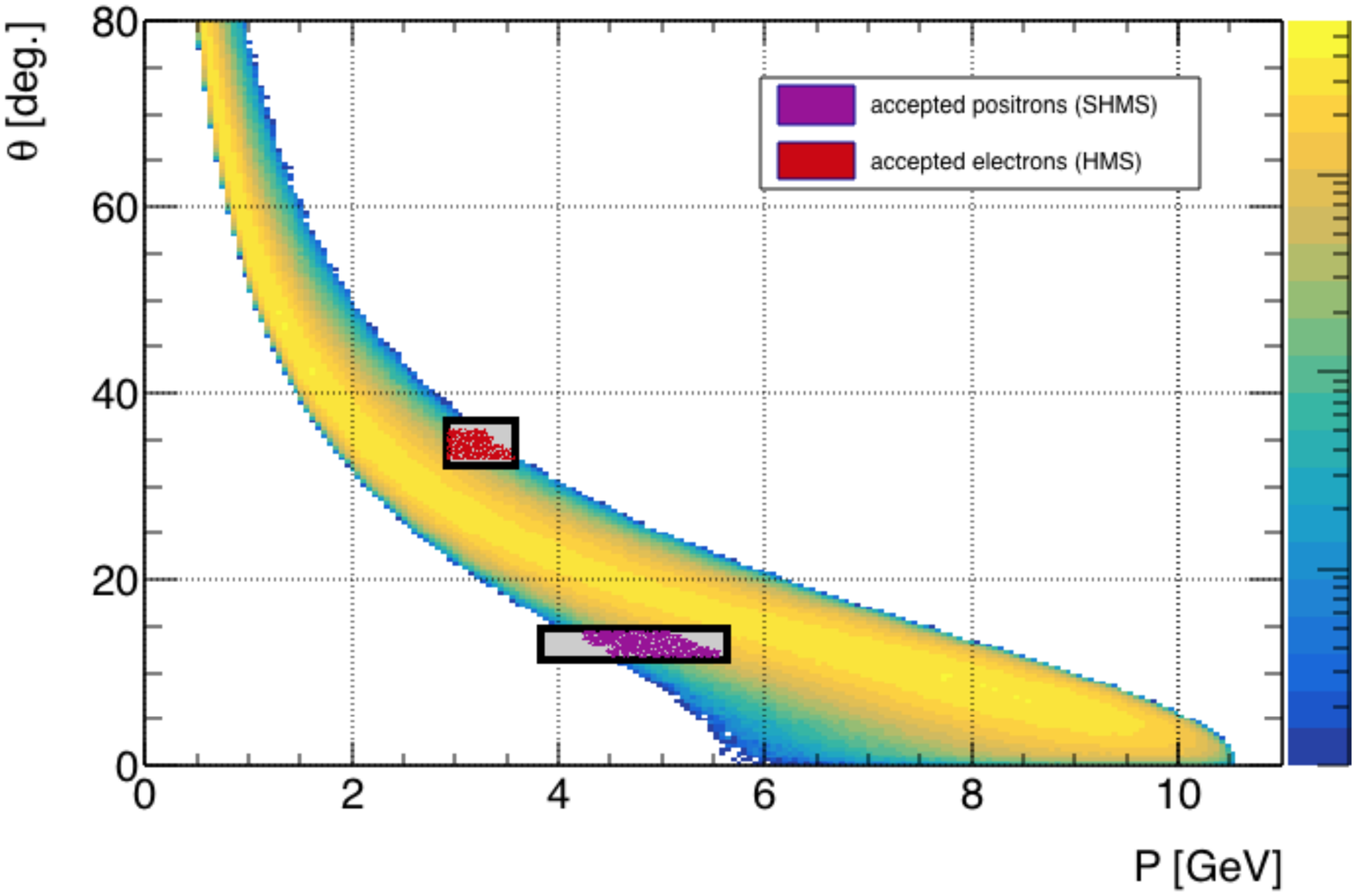}
\includegraphics[scale=1.0,angle=-90,width=0.47\textwidth]{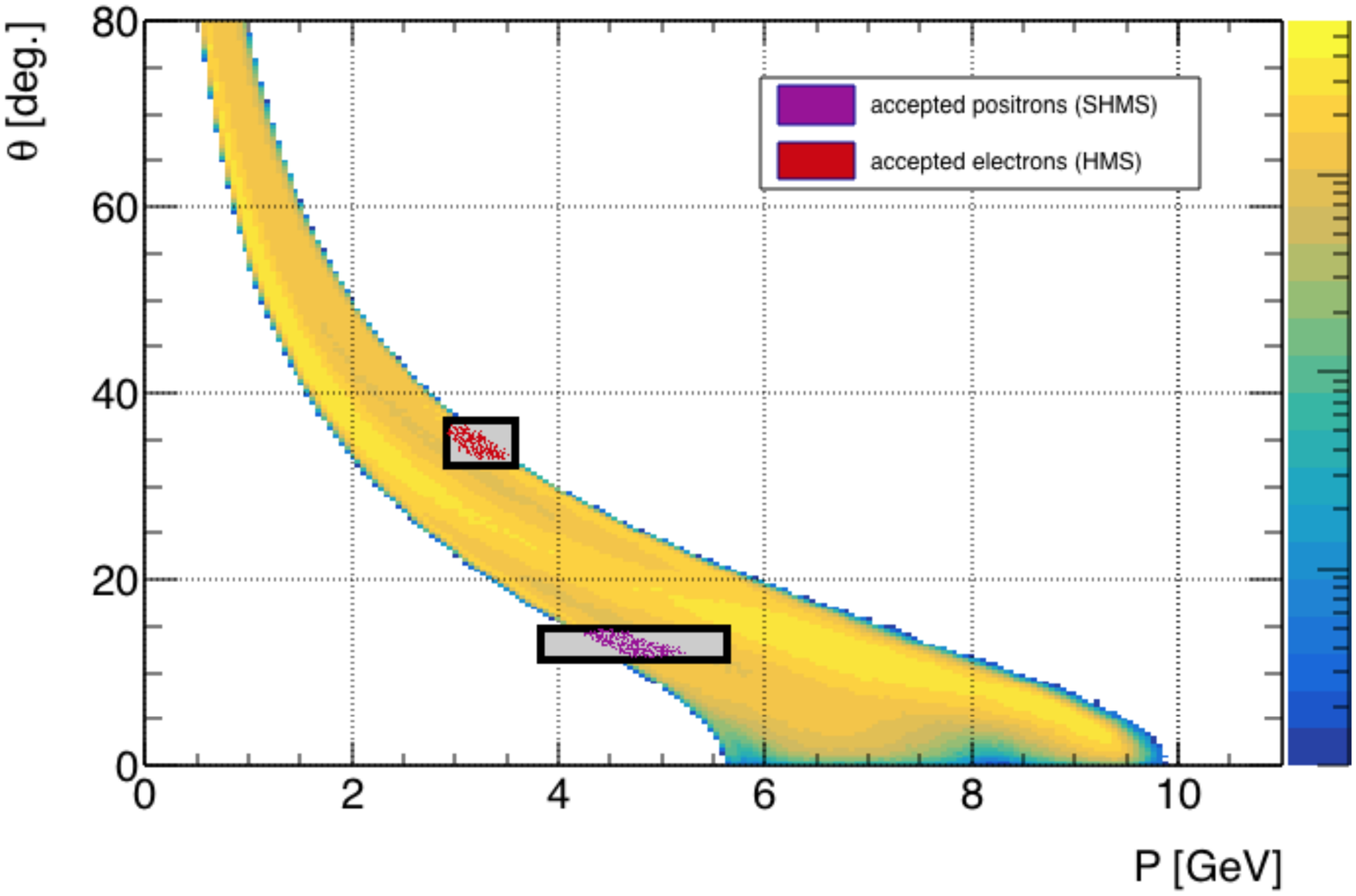}
\caption{Spectrometer settings (rectangular boxes) from the optimization to select the high-$t$
region where the $t$-channel production is highly suppressed compared to the
$P_c$ production rate. See also Fig.~\ref{fig:kin-theta}}
\label{fig:kin-t}
\end{center}
\end{figure}

Shown in Fig.~\ref{fig:kin-t} left and right are the distributions of angle
versus momentum of the decay pair of leptons, the full correlated phase space of
the $t$-channel production of the pair is shown on the left figure while the
similar phase space of the resonant $P_c$ production is shown on the right. 

\begin{figure}[!ht]
\begin{center}
\includegraphics[scale=1.0,angle=0,width=0.6\textwidth]{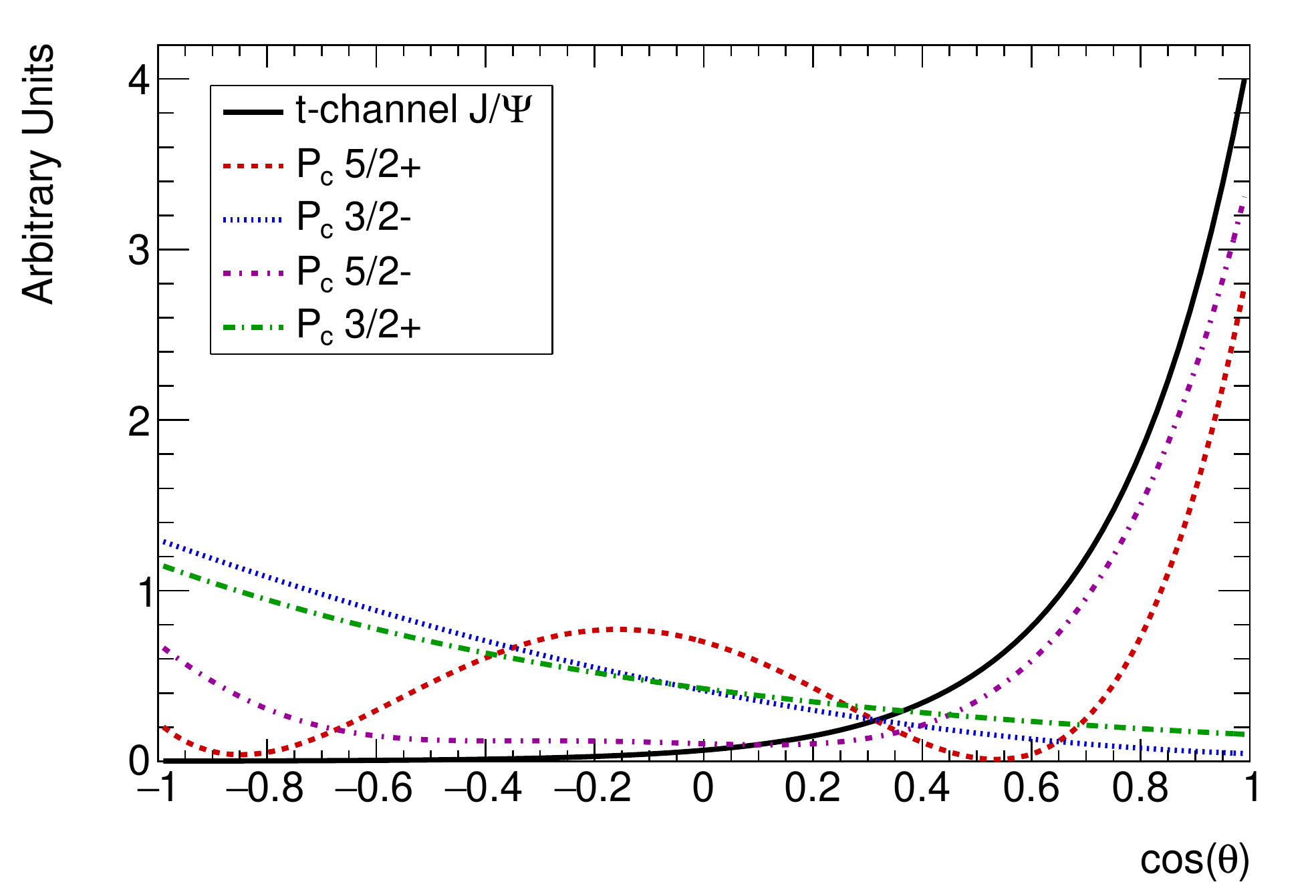}
\caption{Angular distribution of the \jpsi production for the $t$-channel,
 normalized to the same area for each curve in arbitrary units, in comparison with the $J/\psi$ production through the exotic $P_c$ resonant state with various possible spin/parity assumptions. The angle $\theta$ is the relative angle between the J/psi and the photon in the center of mass. Note that, for the $t$-channel this is directly related to the $t$-dependence of the cross section.}
\label{fig:kin-theta}
\end{center}
\end{figure}

It is important to take into account the different possibilities of spin of the
exotic charmed resonance. In all cases we found that it is best to keep
kinematics that correspond to $\cos{\theta}$ between $-0.4$ and $0.2$ as shown
in Fig.~\ref{fig:kin-theta}, or, in other words, close to the 90$^{\circ}$ range in
the center of mass frame. This maximizes the $P_c$ production rate, relative to
the $t$-channel production rate.

\begin{figure}[!ht]
\begin{center}
\includegraphics[scale=1.0,angle=0,width=0.47\textwidth]{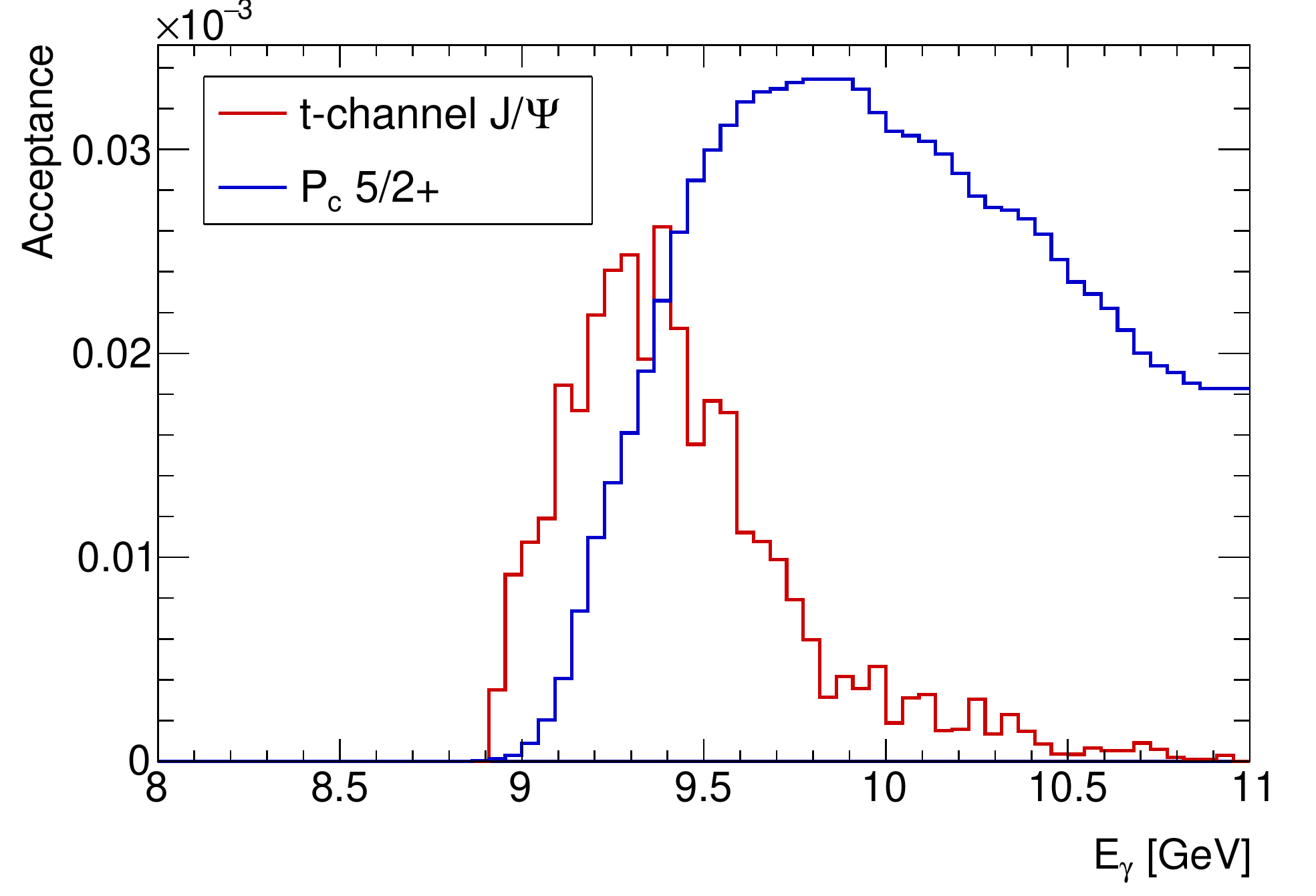}
\includegraphics[scale=1.0,angle=0,width=0.47\textwidth]{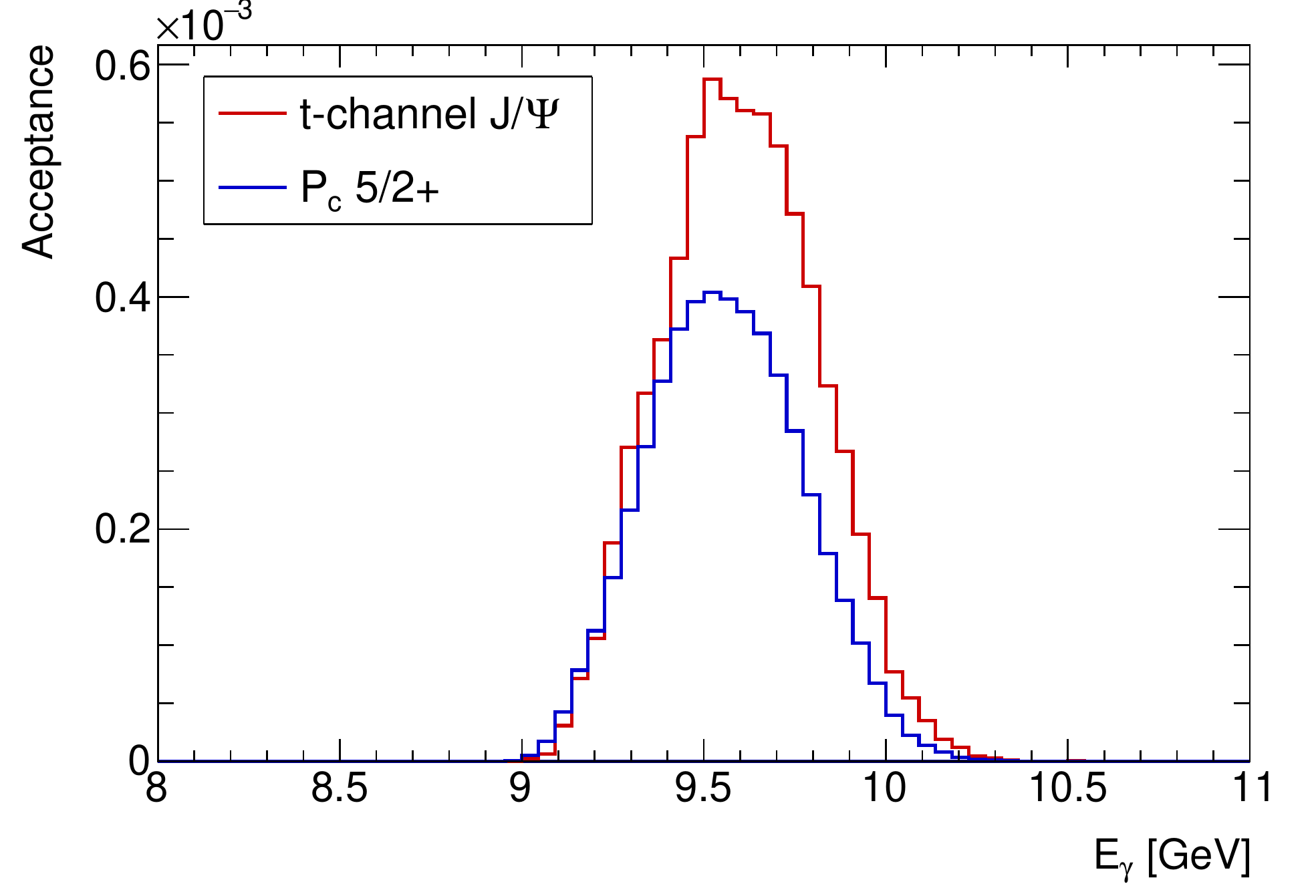}
\caption{The acceptance for setting \#1 (left) and \#2 (right) from
Tab.~\ref{table:kin} as a function of the photon energy $E_\gamma$. The
$t$-channel is shown in red, and the \pc is shown in blue (the angular decay
distribution was taken to be consistent with the 5/2+ assumption shown in
Fig.~\ref{fig:kin-theta}). The acceptance for setting \#1
is perfectly optimized to measure the \pc.}
\label{fig:acc-e}
\end{center}
\end{figure}

\section{Physics and accidental backgrounds}

\begin{figure}[H]
\begin{center}
\includegraphics[scale=1.0,angle=0,width=0.5\textwidth]{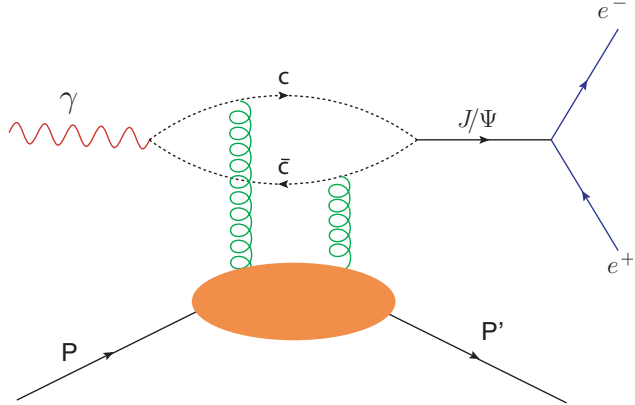}
\caption{t-channel 2-gluon exchange elastic $J/\psi$ photo-production mechanism.}
\label{fig:t-chan-diag}
\end{center}
\end{figure}

The typical process of elastic $J/\psi$ production is usually described by a
Feynman diagram represented in Fig.~\ref{fig:t-chan-diag} and is well understood
at high energies using perturbation theory~\cite{Donnachie:2002en}. This process
at threshold is usually described by a two-gluon
exchange~\cite{Brodsky:2000zc,frankfurt:2002} although at threshold the gluons
could be an effective representation of an interaction that conserves color but
is much more complicated. A full experimental physics program to explore the
threshold region of $J/\psi$ production completely is planned by gathering large
amount of data through electro-production and photo-production using the CLAS12
~\cite{CLAS12-tcs:proposal} and SoLID~\cite{SoLIDjpsi:proposal} detectors in the coming years.

\subsection{Production of $e^+e^-$ through elastic $J/\psi$ production and decay}

\begin{figure}[H]
\begin{center}
\includegraphics[scale=1.0,angle=0,width=0.75\textwidth]{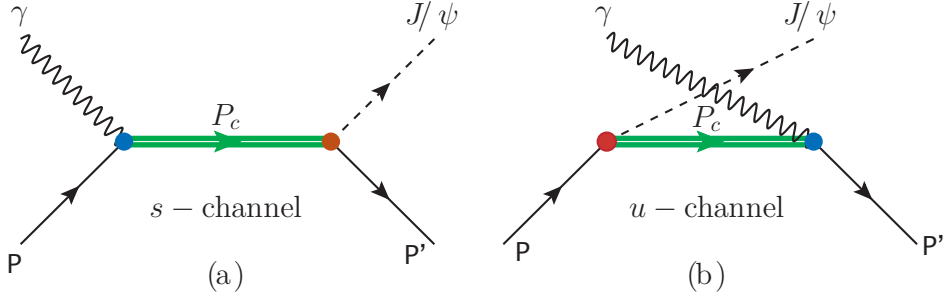}
\caption{$s-$ (a) and $u-$ (b) channel resonant production of  $J/\psi$ through $P_c$.}
\label{fig:s-u-chan-diag}
\end{center}
\end{figure}

In the proposed experiment we are not concerned with the elastic $t$-channel
production of $J/\psi$, which we consider to be a physics background, but are rather
interested in confirming the possible resonance production of the $J/\psi$
through the decay of the newly discovered states at LHCb, namely $P_c(4450)$ and
$P_c(4380)$. This production is typically described by an $s$- and $u$-channel
production of these resonances according to the diagrams of
Fig.~\ref{fig:s-u-chan-diag}. More specifically it is a search for $P_c(4450)$
that we are focused on in this proposal. $P_c(4380)$ is broader with a lower
cross section, and thus requires a more challenging setup to be determined
cleanly.

Therefore, in this experiment the challenge is to separate the two different
processes, one that we consider to be a physics background ($t$-channel production
of $J/\psi$) and one that is our important signal ($s$- and $u$-channel resonant production through $P_c$).
We propose to use spectrometer settings that will dramatically reduce the
acceptance of the $t$-channel production of the \jpsi relative to the  $s$- and
$u$-channel resonant production of \pc. These setting were optimized using a
multidimensional scan of the acceptance for both spectrometers in the full phase
space.

\subsection{Bethe-Heitler pair production}

\begin{figure}[H]
\begin{center}
\includegraphics[scale=0.4,angle=0,width=0.70\textwidth]{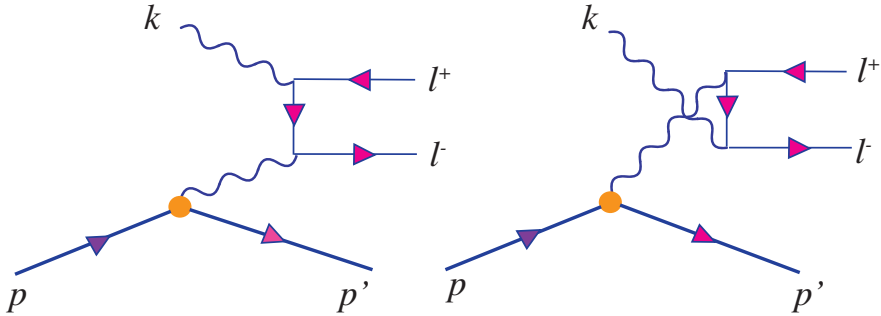}
\caption{Bethe-Heitler (BH) mechanism producing a background process to the
$t$-channel and $P_c$ resonant production.}
\label{fig:bh-t}
\end{center}
\end{figure}

\begin{figure}[H]
\begin{center}
\includegraphics[scale=0.5,angle=0,width=0.50\textwidth]{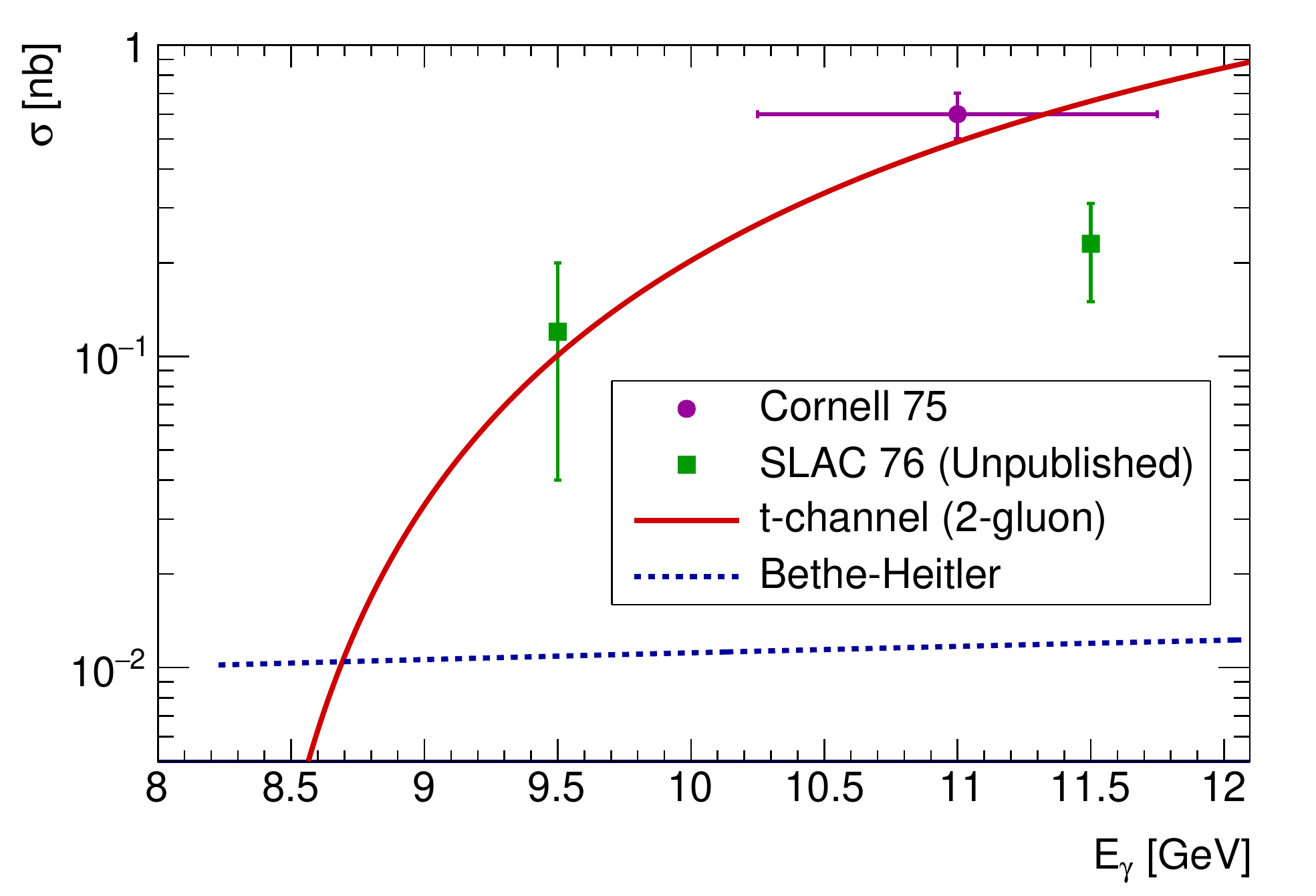}
\caption{B-H rate relative to the elastic \jpsi production  in the $t$-channel.
}
\label{fig:bh-rate}
\end{center}
\end{figure}

To evaluate the Bethe-Heitler background represented by the processes described
in Fig.~\ref{fig:bh-t} we used the calculations of Pauk and
Vanderhaeghen~\cite{Pauk:2015oaa,Vander:2016} but with $M_{l^+l^-}$ evaluated
within the acceptance of our spectrometer settings centered around the mass of
$P_c (4450)$. We use the dipole electromagnetic form factor for the range of
momentum transfers covering the proposed experiment. We find that this
background is over 10 times smaller as shown in Fig.~\ref{fig:bh-rate},
nevertheless it can be calculated and controlled for.

\subsection{Single $e^\pm$ background}
The electron rate in the HMS was estimated using CTEQ5~\cite{Lai:2000im}, and
cross checked using the F1F209 program~\cite{Bosted:2012qc}. The positron rate in
the SHMS was estimated using the EPC program~\cite{Lightbody:1988ke} combined
with a positron background program written for the E94-010 experiment at JLab.
The rate can be found in Tab.~\ref{table:singles}.

\subsection{Single $\pi^\pm$ background}
The charged pion singles rate was estimated using the Wiser program~\cite{wiser:1977}.
Its results can be found in Tab.~\ref{table:singles}.

\subsection{Accidental coincidence rate}
Using the results from Tab.~\ref{table:singles}, the accidental coincidence rate
for a $50\,\text{ns}$ trigger window
between the HMS and SHMS, was found to be of the order of $10^{-5}\,\text{Hz}$.
This is two full orders of
magnitude lower than the expected signal rate and therefore negligible.
For this calculation we assumed a pion rejection larger than 10$^3$ from the
combined \u{C}erenkov and calorimeter system.

\begin{table}[H]
\caption{Singles rates}
\begin{center}
\begin{tabular}{c|cc|cc}
      &\multicolumn{2}{c|}{HMS} & \multicolumn{2}{c}{SHMS}\\
      \hline
      Setting&$e^-$ (kHz)  & $\pi^-$ (kHz) & $e^+$ (kHz) & $\pi^+$ (kHz)\\
      \hline
      \#1 & $6.9\times 10^{-3}$ & $7.5\times 10^{-2}$  
          & $6.5\times 10^{-4}$ & $1.95\times 10^2$ \\
      \#2 & $9.7\times 10^{-1}$ & $2.2\times 10^{0}$  
          & $7.5\times 10^{-4}$ & $10.5\times 10^0$ 
\end{tabular}
\end{center}
\label{table:singles}
\end{table}%

\subsection{Background from $\gamma p\rightarrow J/\psi p \pi $}
The inelastic channel of \jpsi production, where an additional final state pion is produced
but not detected, might contaminate the kinematic region where the \pc is
produced. 
The cross section of these inelastic channels was found to be less than 30\% of
the elastic $t$-channel cross section at high energies, and is expected to be
even smaller near the \jpsi threshold \cite{Binkley:1981kv,Chudakov:2007}.
In this region, the dominant contribution is the resonant channel $\gamma
p\rightarrow \jpsi \Delta(1320)$, with a threshold at approximately
$E_\gamma>9\,\text{GeV}$.

In the acceptance of our proposed setting \#1, and for a photon endpoint energy
of $10.66\,\text{GeV}$, the photon energy spectrum for this
inelastic $t$-channel process occupies the same reconstructed energy range as
the elastic $t$-channel events. These reconstructed energies corresponds to a true photon energies of
$1\,\text{GeV}$ higher, where the cross is approximately four times larger. This
fourfold rise in the cross section is almost exactly compensated by the acceptance, which is
four times lower for this process, due to a corresponding shift of the
reconstructed $t$ of $1\,\text{GeV}^2$. Ultimately, this background is expected to
increase the elastic $t$-channel background by at most 30\%.

\subsection{Background from lepto-production}
To estimate the background due to lepto-production, we simulated the $e p
\rightarrow e \gamma^* p \rightarrow e \jpsi p$ process for a $50\,\mu\text{A}$
electron beam at $11\,\text{GeV}$. We found only quasi-real photons, up to
a virtuality of $Q^2 ~ 0.01\,\text{GeV}^2$ to have a significant impact.  
Photons with a higher virtuality are highly suppressed because of the following
reasons:
\begin{itemize}
\item the virtual photon flux drops with $Q^2$
\item Higher $Q^2$ means lower $W^2$ for fixed values of $\nu$, and the
$t$-channel cross section drops for lower $W^2$. Furthermore, close to
threshold, the available phase space shrinks rapidly for lower $W^2$.
\item The highly-tuned spectrometer acceptance drops with $Q^2$.
\end{itemize}
Because only the quasi-real photons play a role, the contribution due to
lepto-production will lead to a (small) enhancement of the count rates.
We will verify the impact of this contribution by conducting a dedicated
measurement without the radiator.

\section{Simulation of the experiment}

We use a custom Monte-Carlo generator to obtain a realistic estimate of the
\jpsi photo-production rates. This generator uses the bremsstrahlung spectrum for a
10\% radiator, appropriate models for the $t$-channel and $P_c$ resonant
channel, and the HMS and SHMS spectrometer acceptance with realistic smearing
effects. The leptonic $\jpsi\rightarrow e^+e^-$ decay is simulated using a
$(1+cos^2\theta_{e})$ angular distribution in the \jpsi helicity frame.
More details about the simulation can be found below.

\subsection{Model for the $t$-channel cross section}
In order to calculate the cross section for the $t$-channel production, we fit
the cross section ansatz for two gluon exchange from Brodksy et al.~\cite{Brodsky:2000zc} (equation (3))
to the available world data. The result of this fit (in
$\text{nb}/\text{GeV}^2)$ is given by,
\begin{align}
\frac{d\sigma}{dt} = 
A v \frac{(1-x)^2}{M_{\jpsi}^2}(s-M_p^2)^2
\exp{bt},
\label{eq:tchannel-xsec}
\end{align}
where $x$ is given by a near-threshold definition of the fractional momentum
carried by the valence quark, $v$ is a kinematic factor,
$b$ the impact parameter and $A$ an overall normalization constant that was
determined by a fit to the world data,
\begin{align}
x&=\frac{2M_{\jpsi}M_p +M_{\jpsi}^2}{s-M_p^2}\\
v&=\frac{1}{16\pi(s-M_p^2)^2}\\
b&=1.13\,\text{GeV}^{-2},\\
A&=6.499\times10^3\,\text{nb}.
\end{align}
Additionally, $M_p$ and $M_{\jpsi}$ are respectively the proton mass and \jpsi mass in
GeV.
The curve from Eq.~\ref{eq:tchannel-xsec} is shown as a red line in
Figures~\ref{fig:xsection}~and~\ref{fig:xsec-comp}. 

\subsection{Model for the $P_c\rightarrow\jpsi p$ cross section}

\begin{figure}[!ht]
\begin{center}
\includegraphics[scale=1.0,angle=0,width=0.5\textwidth]{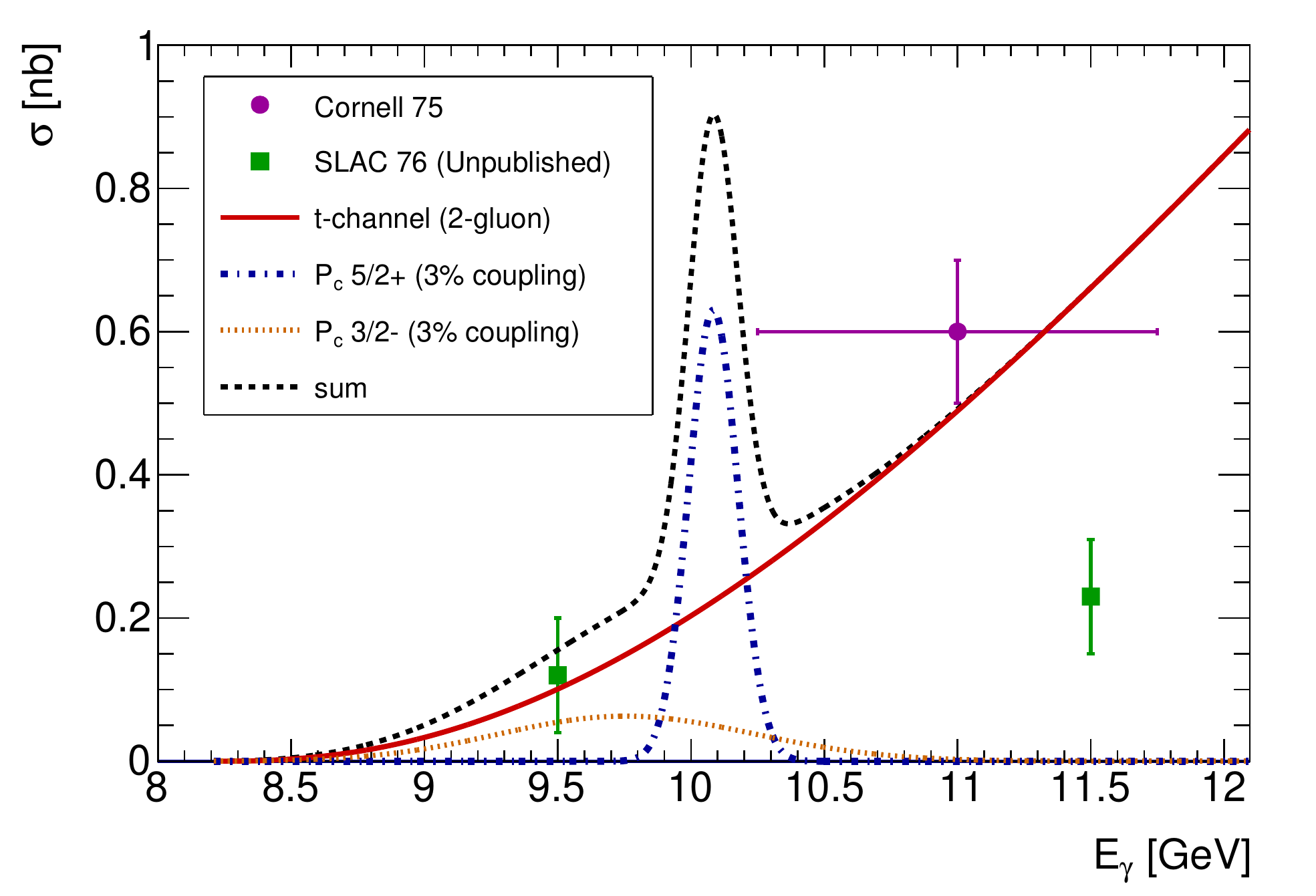}
\caption{$J/\psi$ production cross section as a function of the photon energy.
The $P_c$ resonant production is shown for the (5/2+,~3/2-) case assuming 3\%
coupling, compared with the available measurements in this region
\cite{Gittelman:1975ix,Anderson:1976sd}.}
\label{fig:xsec-comp}
\end{center}
\end{figure}

Several equivallent approaches to calculate the 
$\gamma p\rightarrow P_c\rightarrow\jpsi p$ cross section can be found in the
literature \cite{Wang:2015jsa,Karliner:2015voa,Kubarovsky:2015aaa,Blin:2016dlf}.
We based our model of the cross section on the work by Wang et al.~\cite{Wang:2015jsa}.
Note that this cross section depends quadratically on the coupling to the 
$\jpsi p$ channel.
We considered the (5/2+) and (5/2-) spin/parity assumptions for the narrow 
\pc state, with the corresponding (3/2-) and (3/2+) assumption for the \pcwide
state. The angular distribution for the \jpsi production for each of the
spin-parity assumptions can be found in Fig.~\ref{fig:kin-theta}.
The contributions of the (5/2+,3/2-) channels to the \jpsi photo-production cross
section as a function of photon energy $E_\gamma$ are shown in
Fig.~\ref{fig:xsec-comp}.

We optimized the spectrometer settings for a (5/2+) \pc case with 5\%
coupling to the $\jpsi p$ channel, as it agrees well with the existing
photo-production data. This setting, also has a good sensitivity to a (5/2-) \pc
as its production cross section is a full order of magnitude larger.
To perform this optimization, a total of 3.4 million possible spectrometer
settings were considered. We selected a setting that maximizes the acceptance for
\jpsi produced with a $\cos\theta$ between $-0.4$ and $0.2$ in the center-of-mass
frame, as shown in Fig.\ref{fig:kin-theta}. This corresponds to a setting that
selects the high-$t$ region, where there is a maximum sensitivity to the (5/2+)
\pc resonant production, while simultaneously the sensitivity to the
$t$-channel \jpsi production is highly suppressed. This setting is listed on the
first line of Tab.~\ref{table:kin}.

\subsection{Bremsstrahlung spectrum}

The generator uses equation (24) from
Tsai~\cite{Tsai:1966js} to evaluate for the bremsstrahlung spectrum. For a 9\%
radiator combined with 1\% from the target (a total of 10\% radiator), the photon beam has an integrated intensity of 2.3\% of the primary electron beam.

\subsection{Detector acceptance and resolution}

\begin{figure}[H]
\begin{center}
\includegraphics[scale=1.0,angle=0,width=0.5\textwidth]{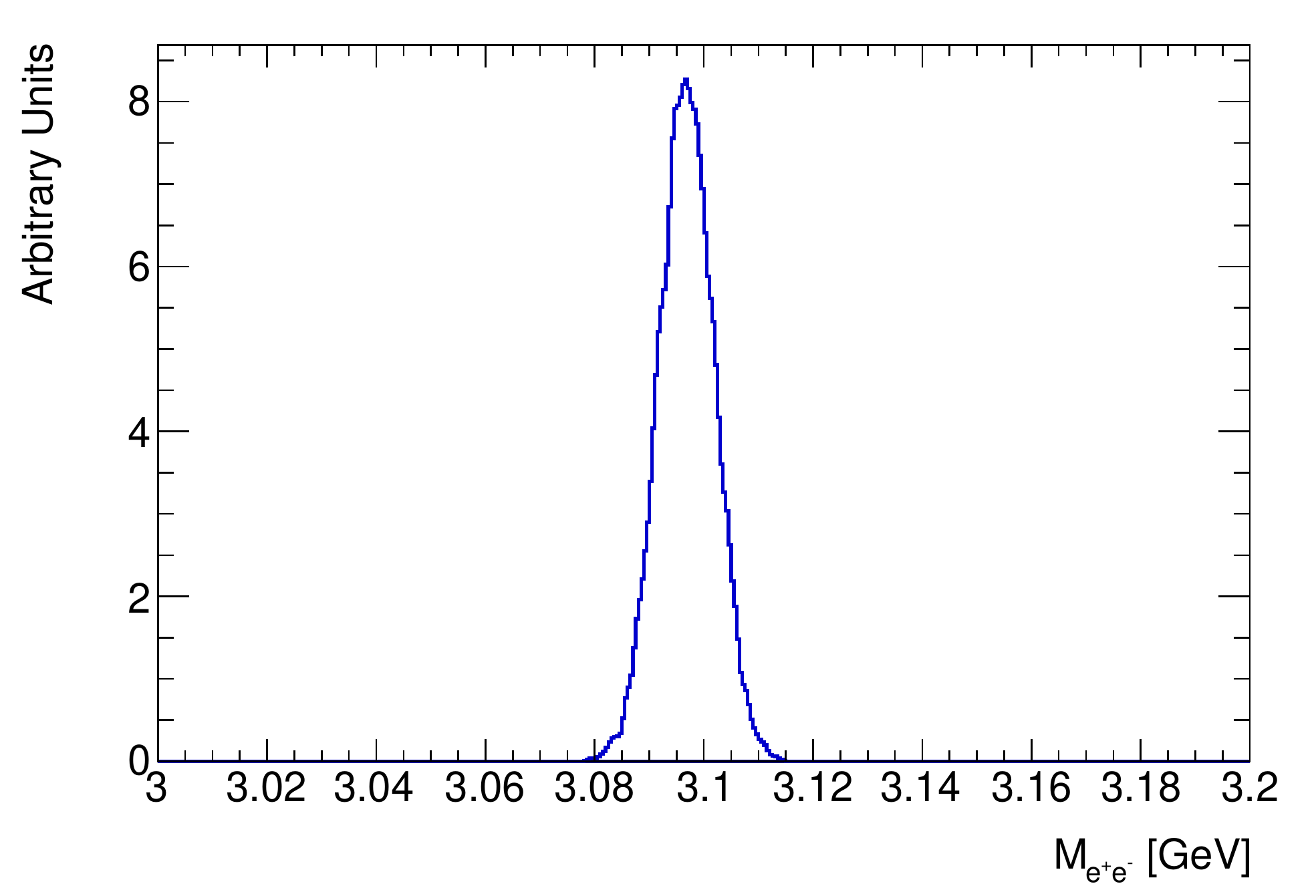}
\caption{
  Invariant mass of the detected lepton pair with realistic smearing. The
  invariant mass resolution is $5\,\text{MeV}$.
}
\label{fig:invmass}
\end{center}
\end{figure}

The spectrometer acceptance and realistic smearing are simulated using the
parameters listed in Tab.~\ref{table:spec}. 
An $e^+e^-$ invariant mass spectrum that was generated using the optimized
setting listed on the first line of Tab.~\ref{table:kin} can be found in
Figure~\ref{fig:invmass}. The reconstructed \jpsi mass resolution is
$5\,\text{MeV}$.

\begin{table}[ht]
\caption{Properties of the Hall C spectrometers.}
\begin{center}
\begin{tabular}{cccccccccc}
     & $P$ & $\Delta P/P$ & $\sigma P/P$ & $\theta^\text{in}$ &
     $\Delta\theta^\text{in}$ & $\Delta\theta^\text{out}$ &
     $\Delta\Omega$ & $\sigma\theta^\text{in}$ &
     $\sigma\theta^\text{out}$ \\
     & GeV/$c$ & \%& \%& & mrad & mrad& msr & mrad & mrad \\
     \hline\\
     HMS & 0.4-7.4 & -10 +10 & 0.1 & 10.5$^\circ$-90$^\circ$ & $\pm 24$ &
     $\pm 70$ & 8 & 0.8 & 1.0 \\
     SHMS & 2.5-11. & -15 +25 & 0.1 & 5.5$^\circ$-25$^\circ$ & $\pm 20$ &
     $\pm 50$ & 4 & 1.0 & 1.0 \\
\end{tabular}
\end{center}
\label{table:spec}
\end{table}%

\clearpage
\section{Projected results}
\label{sec:res}

In this section we describe the results of our simulation and the expected
results for 9 days of beam on target. We discuss the projected yields in
case of 5\% coupling for the $\pc\rightarrow J/\psi p$ channel.
Additionally, we will quantify the statistical precision with which we can
identify the $P_c$ resonance for different values of the coupling.
Finally, we will show the estimated impact of this experiment on the available
world data for the \jpsi photo-production cross section.

\subsection{Projected results in case of 5\% coupling}

\begin{figure}[htb]
\begin{center}
\includegraphics[scale=1.0,angle=0,width=0.45\textwidth]{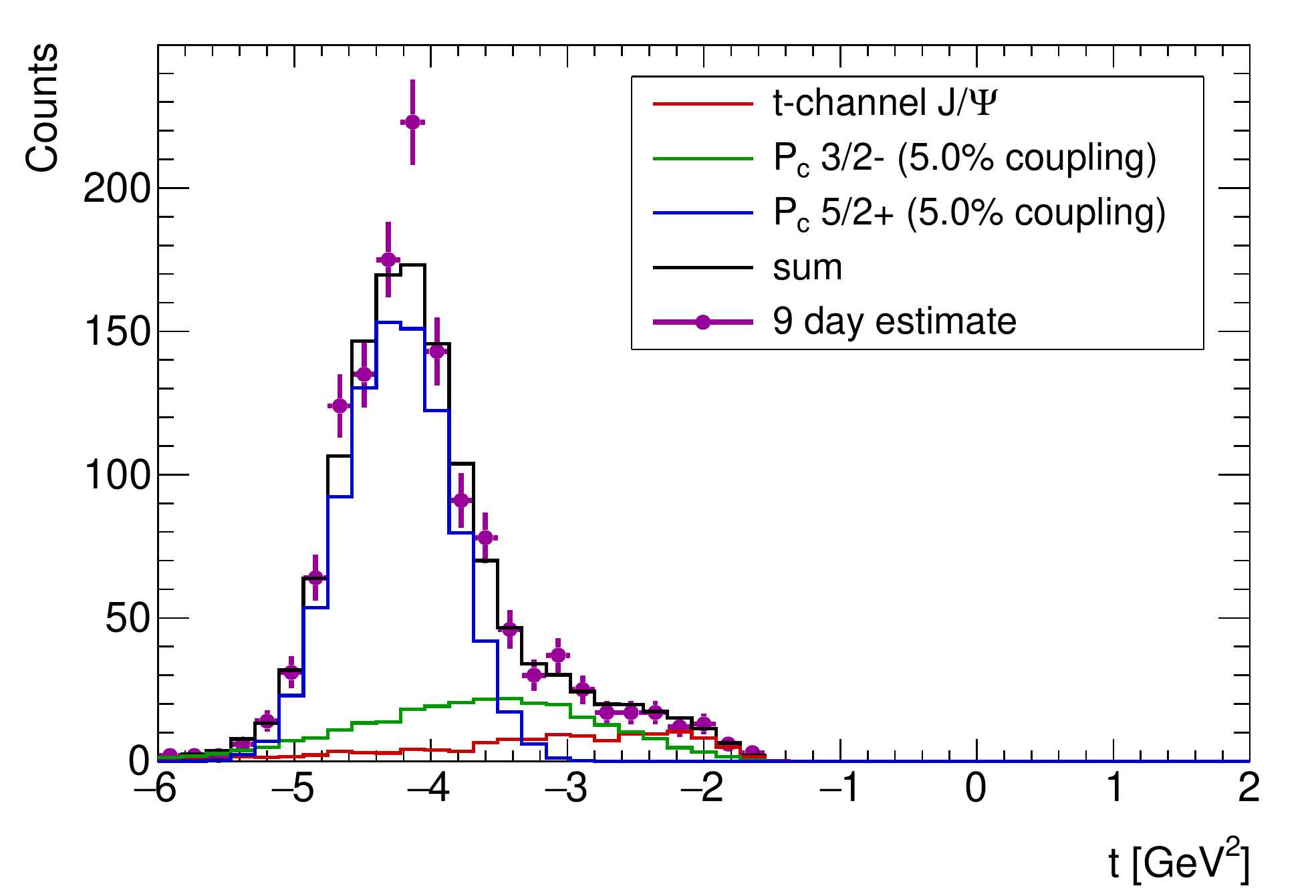}
\includegraphics[scale=1.0,angle=0,width=0.45\textwidth]{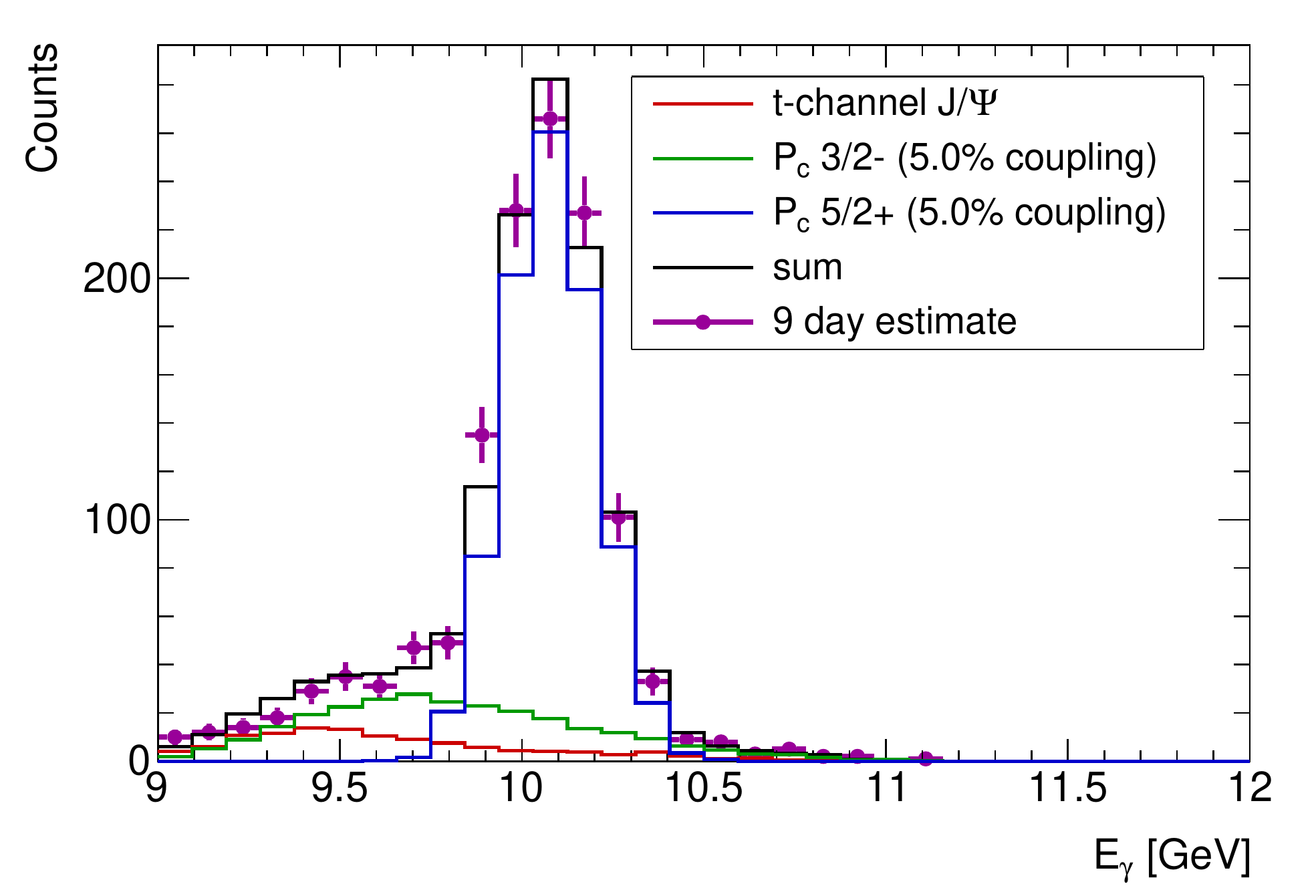}
\caption{Expected results for the reconstructed $t$ and $E_\gamma$ spectrum
for 9 days of beam on target, assuming the most
probable (5/2+,~3/2-) $P_c$ from \cite{Wang:2015jsa} with 5\% coupling. There is
clear separation in both spectra between the $P_c$ (5/2+) resonant channel, and
the $t$-channel.}
\label{fig:yieldone}
\includegraphics[scale=1.0,angle=0,width=0.45\textwidth]{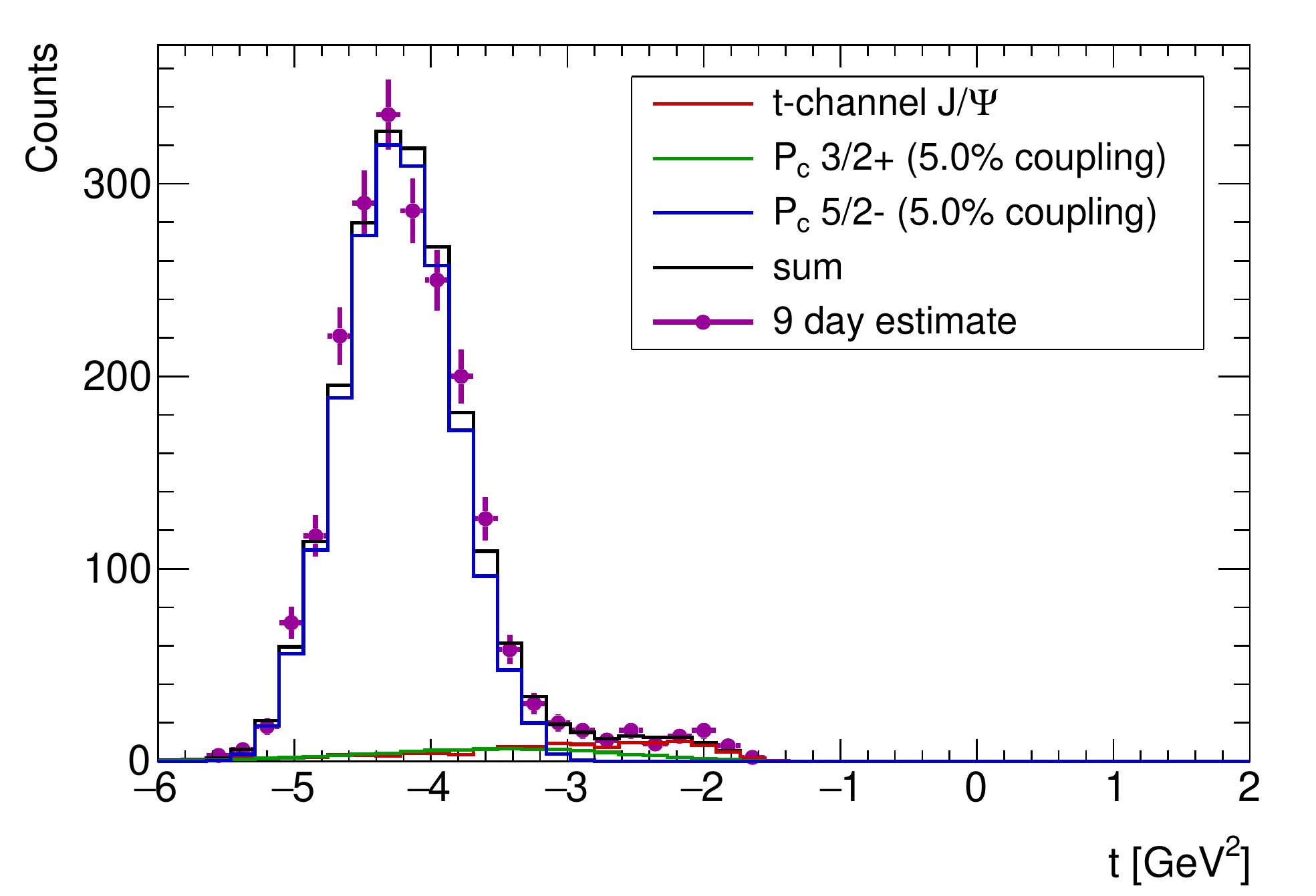}
\includegraphics[scale=1.0,angle=0,width=0.45\textwidth]{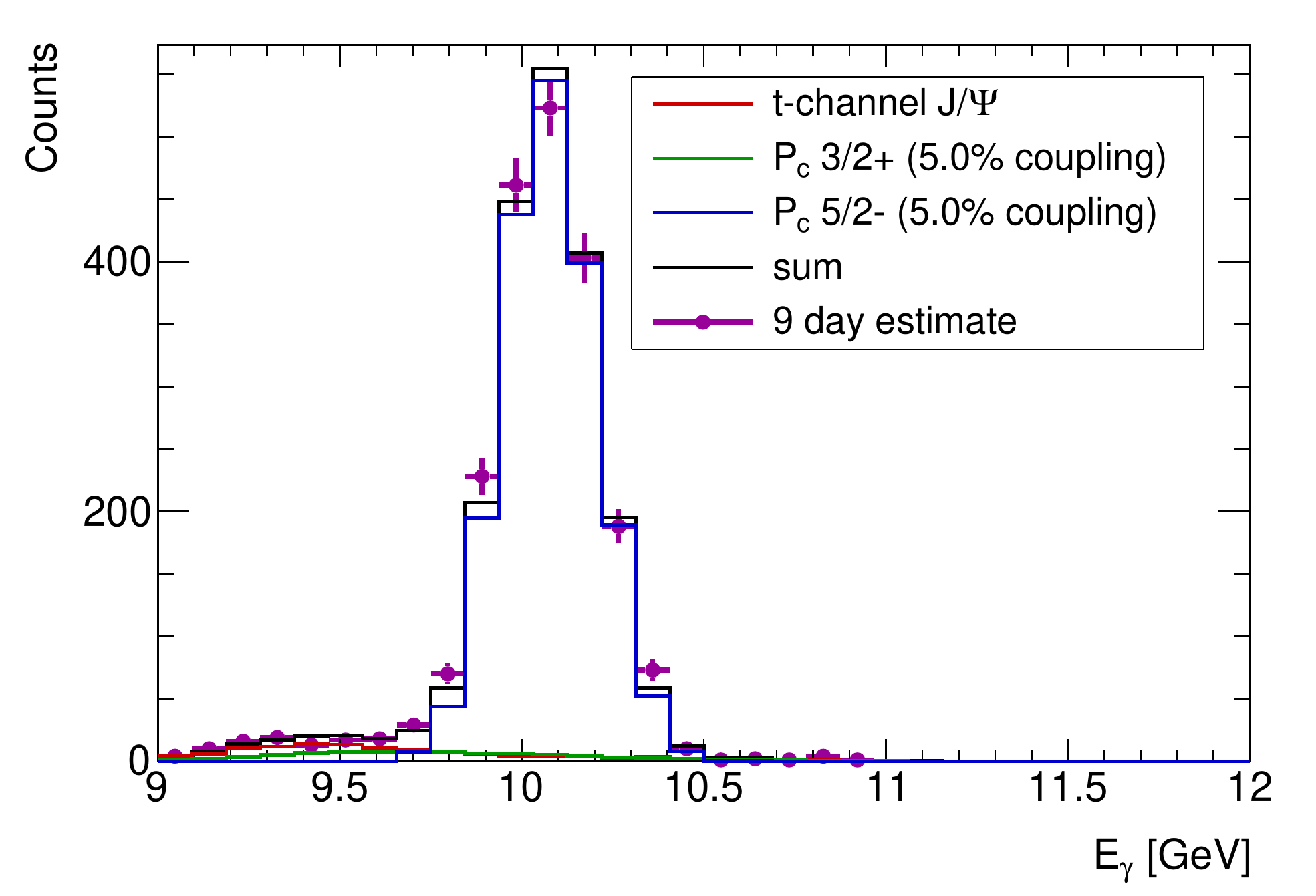}
\caption{Expected results for the reconstructed $t$ and $E_\gamma$ spectrum
for 9 days of beam on target, assuming the less probable
(5/2-,~3/2+) $P_c$ from \cite{Wang:2015jsa} with 5\% coupling. Due to the larger
cross section for the 5/2-, the separation in both spectra is even better than for
the 5/2+ assumption shown in Fig.~\ref{fig:yieldone}.}
\label{fig:yieldtwo}
\end{center}
\end{figure}

As shown in Fig.~\ref{fig:yieldone} and
Fig.~\ref{fig:yieldtwo} the results, clearly reveal
the resonant structure of the pentaquark assuming a 5\% coupling. 

\begin{figure}[htb]
\begin{center}
\includegraphics[scale=1.0,angle=0,width=0.45\textwidth]{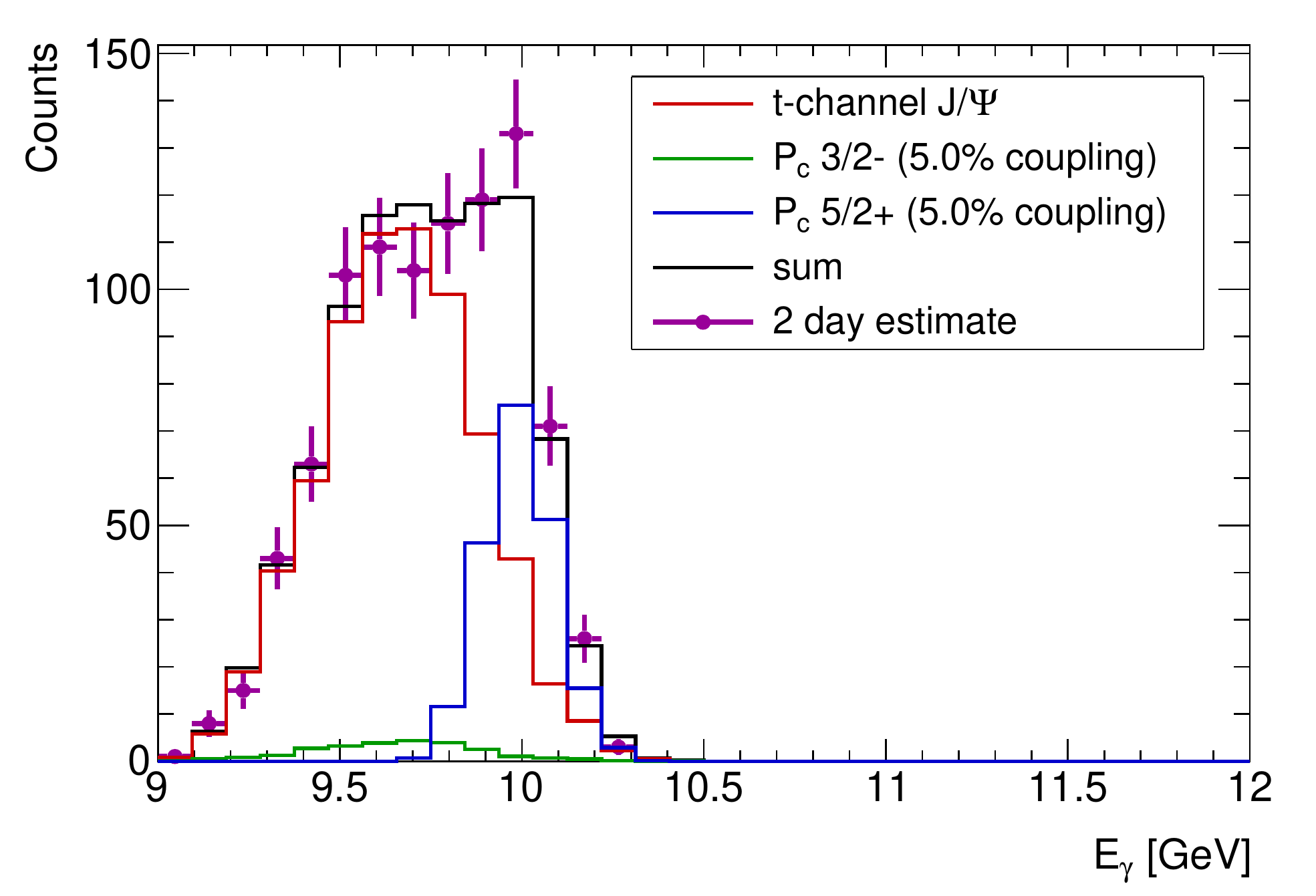}
\includegraphics[scale=1.0,angle=0,width=0.45\textwidth]{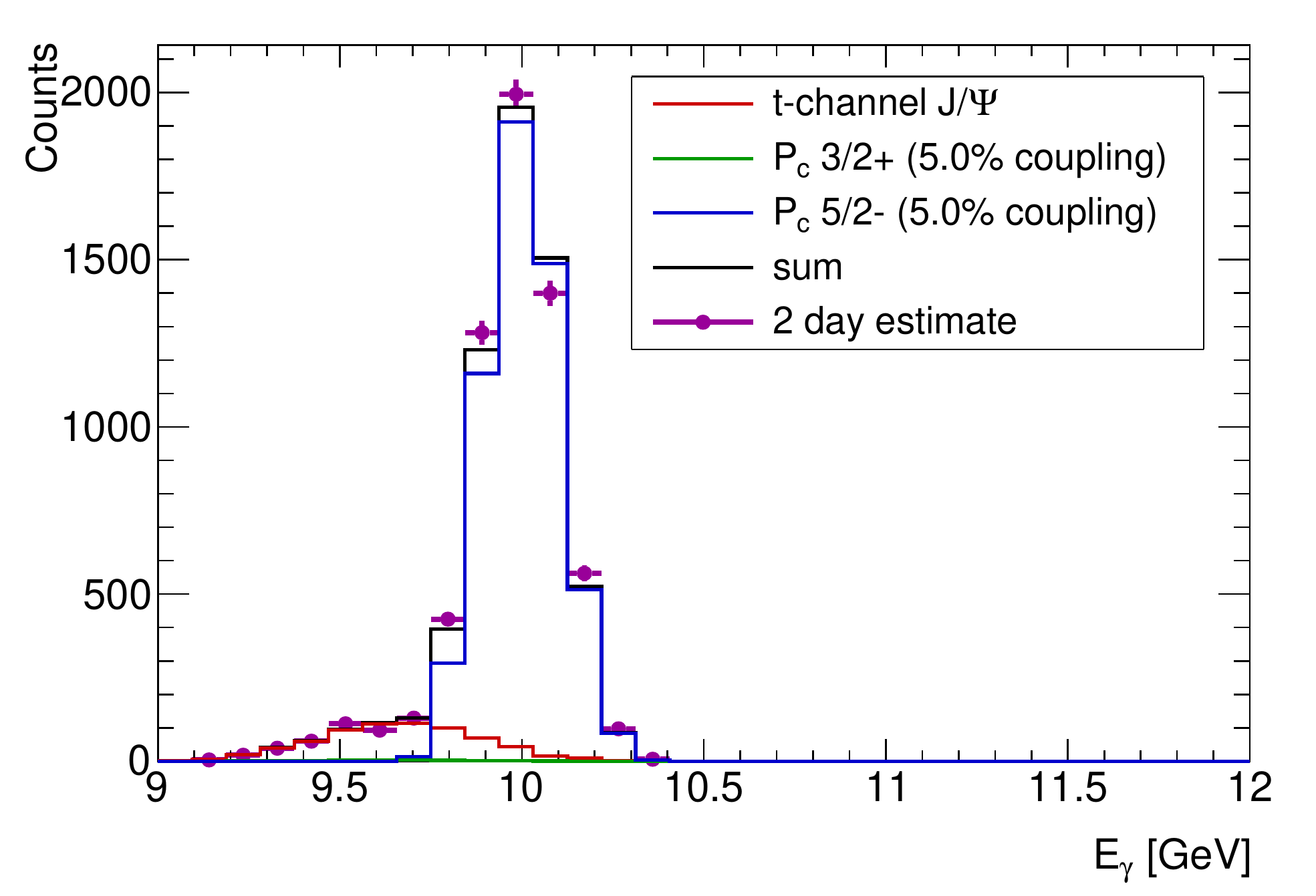}
\caption{Expected results for the reconstructed $E_\gamma$ spectrum
for the calibration measurement with 2 days of beam on target. The left panel
  shows the (5/2+,~3/2-) case , and the right panel shows the
  (5/2-,~3/2+) case, both with 5\% coupling.
}
\label{fig:yieldbg}
\end{center}
\end{figure}

The projected results from the calibration measurement of the $t$-channel \jpsi
background to the $P_c$ resonant channel, for 2 days of beam on target, can be found in
Fig.~\ref{fig:yieldbg}. Note that in addition to providing the necessary
leverage for the $t$-channel background subtraction, 
this calibration measurement will greatly impact our
knowledge of the $t$-channel \jpsi photo-production near threshold, where
currently no world data exist. 

\subsection{Sensitivity to the $P_c$ resonant production}

To obtain an estimate of the sensitivity to the $P_c$ resonant process as a
function of the coupling to the $\jpsi p$ channel, we calculated the
log-likelihood difference $\Delta\log\mathcal{L}$ between the hypothesis that
the
simulated spectra can be described by just a $t$-channel process, and the
hypothesis that the $P_c$ resonances are present on top of the $t$-channel
production. 
We assumed 9 days of beam at $50\,\mu A$ for setting \#1.
We then used Wilk's
theorem~\cite{Wilks:1938dza} to relate the value of
$2\Delta\log\mathcal{L}$ to a value of $\chi^2$ with 5 degrees of freedom (one
for the coupling, and 4 for the mass and width of each of the $P_c$).
Note that a binned likelihood approach was used, which yields a conservative
estimate compared to the results of a full unbinned extended maximum likelihood
procedure.

\begin{figure}[bht]
\begin{center}
\includegraphics[scale=1.0,angle=0,width=0.45\textwidth]{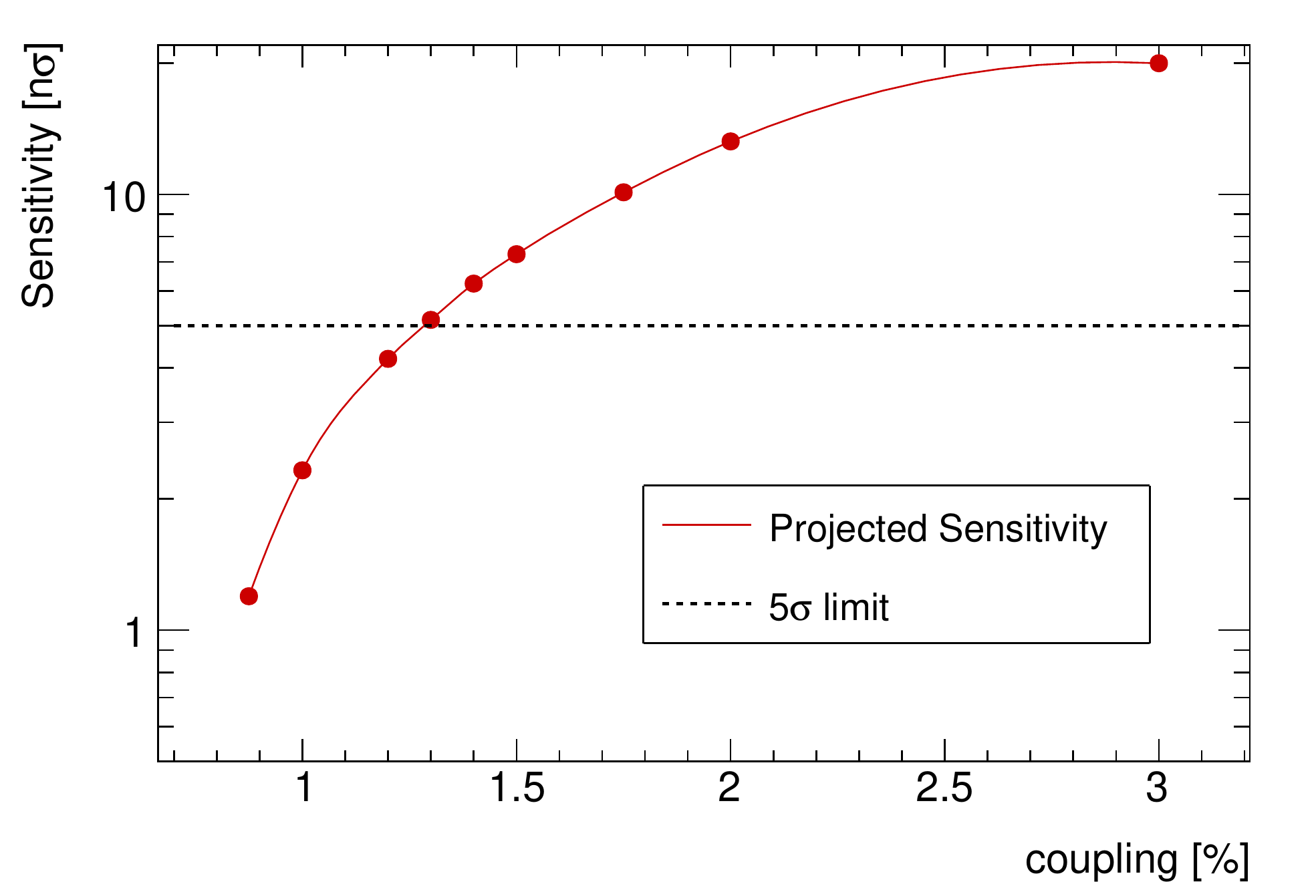}
\includegraphics[scale=1.0,angle=0,width=0.45\textwidth]{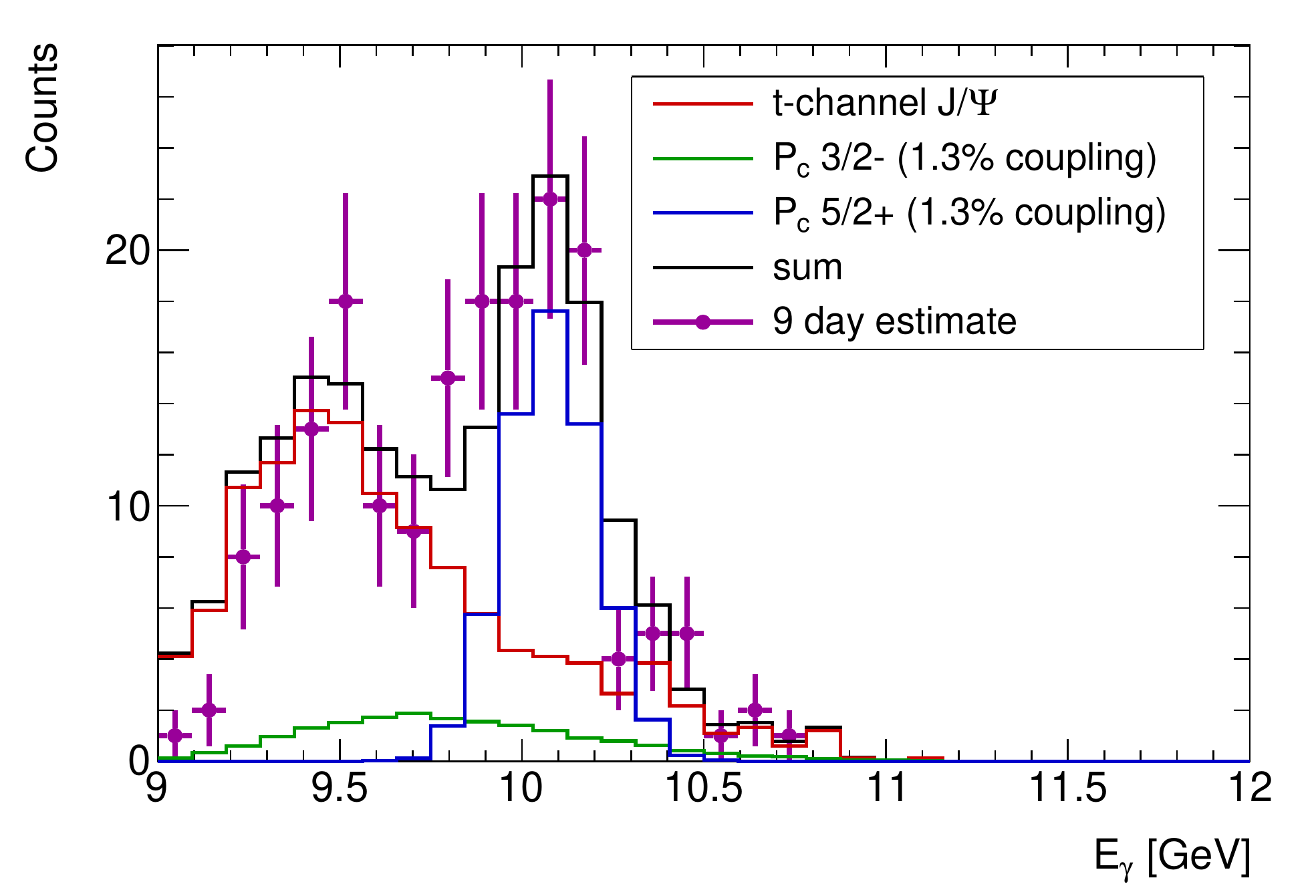}
\caption{
  The left figure shows the sensitivity to the $P_c$ as a function of the
  coupling to the $\jpsi p$ channel, obtained from a log-likelihood analysis.
  The dashed line shows the $5\sigma$ level of sensitivity necessary for
  discovery. This level is reached starting from a coupling of 1.3\%.
  The right panel shows the expected results for the reconstructed $E_\gamma$
  spectrum for this 1.3\% coupling for the \pc (5/2+).}
\label{fig:sensitivity}
\end{center}
\end{figure}

The results of this sensitivity study can be found in
Fig.~\ref{fig:sensitivity}. We found that, for values of the coupling of 1.3\%
and higher, we have a sensitivity of more than the 5 standard deviations for
discovery. Fig.~\ref{fig:sensitivity} also shows the projected results in case
of a 1.3\% coupling. For a coupling of 5\%, our sensitivity far exceeds 
20 standard deviations.

In the proposal, we assumed a realistic coupling of 5\% from
Wang~\cite{Wang:2015jsa}, which they found to be compatible with the currently
existing $\jpsi$ photo-production data.
A more recent statistical analysis by Blin~\cite{Blin:2016dlf} found an upper
limit of the coupling values to be between $8-17\%$ at the 95\% confidence
level for the \pc (5/2+).
Furthermore, Karliner~\cite{Karliner:2015ina} argues that the coupling cannot be
too small, as the $\pc\rightarrow \jpsi p$ signal is 4.1\% of the $\jpsi p$
final state in $\Lambda_b\rightarrow K^-\jpsi p$. If the coupling were too
small, the value of $\Lambda_b\rightarrow K^- P_c$ with the $P_c$ decaying to
final states other than $\jpsi p$, would become unreasonably large in comparison with
the measured branching fraction of $\Lambda_b \rightarrow K^-\jpsi p$.
This means that, due to the sensitivity of the proposed experiment down to very
low values of the coupling, we will have the ability to provide a very strong
exclusion of the charmed-pentaquark assumption in case it is not found.

\subsection{Projected impact on the world data for \jpsi production}

\begin{figure}[th]
\begin{center}
\includegraphics[width=0.80\textwidth]{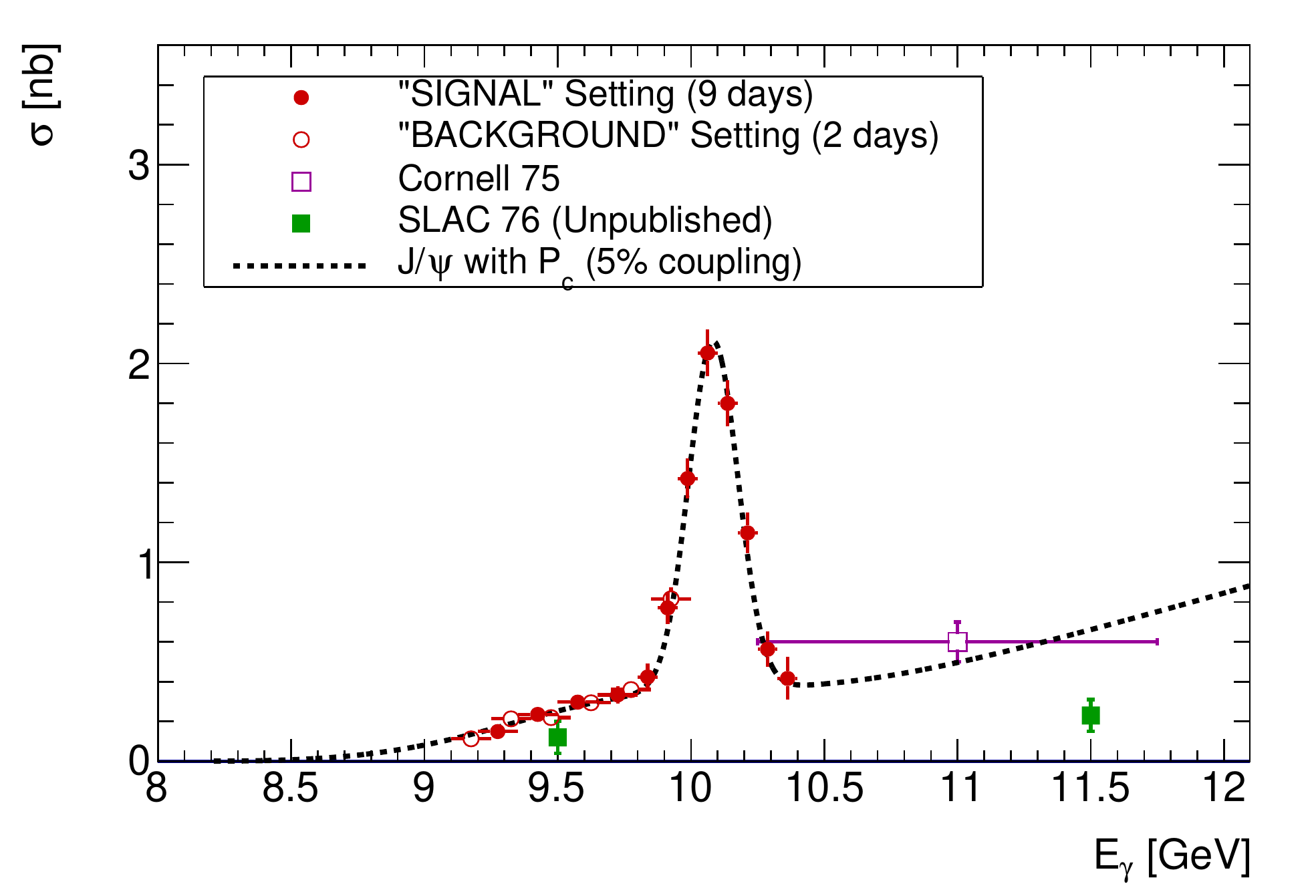}
\caption{
  Projected impact of this experiment assuming, the (5/2+,~3/2-) case with 5\%
  coupling, for 9 days of data taking in setting \#1 (solid circles) and 2
  additional days of data taking in setting \#2 (open circles).
  The existing data points from Cornell data from~\cite{Gittelman:1975ix} and
  SLAC (unpublished)~\cite{Anderson:1976sd} are also shown. 
}
\label{fig:impact}
\end{center}
\end{figure}

The projected impact of the proposed experiment, assuming the (5/2+,~3/2-) case
with 5\% coupling is shown in Fig.~\ref{fig:impact}. These results will
dramatically enhance our knowledge of \jpsi photo-production near threshold.
The absolute cross section measurements from this experiment will provide
valuable input for future experimental endeavors at CLAS12 and SoLID
\cite{CLAS12-tcs:proposal,SoLIDjpsi:proposal}.

\section{Run plan and beam request}
We propose to carry the measurement of elastic photo-production of \jpsi in
the threshold region with the aim to confirm the LHCb $P_c(4450)$ discovery. The
experiment uses the standard equipment of the upgraded Hall C apparatus
at Jefferson Lab. We request 11 days (264 hours) of beam time.
The first 40 hours will focus on measuring the shape of the $t$ distribution
with high statistics, using setting \#2 to maximize the combined acceptance
for this process. We will take an additional 8 hours
of data in this setting without the radiator, in order to assess the
contribution from lepto-production.
Finally, we will conduct our main measurement in setting \#1 for the remaining
216 hours.
See Tab.~\ref{table:kin} for the definitions of the spectrometer settings \#1 and
\#2.
Accidental coincidences  between the two spectrometers will be measured at the same setting
and the same time in the momentum acceptance of the spectrometers outside the
true physics events. 

We request 11 days to perform this high-impact measurement in search of the LHCb
charmed exotic resonances consistent with "pentaquarks".

\clearpage
\bibliography{Pc_main}%

\end{document}